\documentstyle[12pt,epsf]{article}
\newcommand{\beq}{\begin{equation}}
\newcommand{\eeq}{\end{equation}}
\newcommand{\bea}{\begin{eqnarray}}
\newcommand{\eea}{\end{eqnarray}}
\newcommand{\beas}{\begin{eqnarray*}}
\newcommand{\eeas}{\end{eqnarray*}}
\newcommand{\nn}{\nonumber}
\newcommand{\limit}{\rightarrow}
\newcommand{\cN}{{\cal N}}
\newcommand{\tr}{{\rm tr}}
\newcommand{\Tr}{{\rm Tr}}
\newcommand{\bra}{\langle}
\newcommand{\ket}{\rangle}
\newcommand{\cO}{{\cal O}}
\newcommand{\st}{\tilde{\sigma}}
\newcommand{\lt}{\tilde{\lambda}}
\renewcommand{\theequation}{\arabic{section}.\arabic{equation}}
\begin{document}
\topmargin 0pt
\oddsidemargin 5mm
\headheight 0pt
\topskip 0mm
\begin{titlepage}
\begin{flushright}
November 2000 \\
KEK-TH-727 \\
SACLAY-SPHT/T00/165 \\ 
hep-th/0011175
\end{flushright}
\vspace{1cm}
\begin{Large}
\begin{center}
{\sc Gaussian and Mean Field Approximations}\\
{\sc for Reduced Yang-Mills Integrals} \\
\end{center}
\end{Large}
\vspace{5mm}
\begin{center}
\begin{large}
{\sc Satsuki Oda}\footnote{E-mail address: {\tt oda@ccthmail.kek.jp}\label{Oda}} 
{\sc and  Fumihiko Sugino}\footnote{E-mail address: 
{\tt sugino@spht.saclay.cea.fr}\label{Sugino}} \\
\end{large}
\vspace{4mm}
${}^{\ref{Oda}}${\it Institute of Particle and Nuclear Studies,}\\
{\it High Energy Accelerator Research Organization (KEK),}\\
{\it Tsukuba, Ibaraki 305-0801, Japan}\\

\vspace{3mm}
${}^{\ref{Sugino}}${\it Service de Physique Th\'{e}orique, C.E.A. Saclay,}\\
{\it F-91191 Gif-sur-Yvette Cedex, France}\\

\vspace{3cm}
\begin{large} 
Abstract 
\end{large}
\end{center}

   In this paper, we consider bosonic reduced Yang-Mills integrals by using 
some approximation schemes, which are a kind of mean field approximation 
called Gaussian approximation and its improved version. 
We calculate the free energy and the expectation values of various operators 
including Polyakov loop and Wilson loop. 
Our results nicely match to the exact and the numerical results obtained before. 
Quite good scaling behaviors of the Polyakov loop and of the Wilson loop can be seen 
under the 't Hooft like large $N$ limit for the case of the loop length smaller. 
Then, simple analytic expressions for the loops are obtained. 
Furthermore, we compute the Polyakov loop and the Wilson loop for the case of the 
loop length sufficiently large, 
where with respect to the Polyakov loop 
there seems to be no known results in appropriate literatures 
even in numerical calculations. 
The result of the Wilson loop exhibits a strong resemblance to the result simulated 
for a few smaller values of $N$ in the supersymmetric case.

\end{titlepage}


\section{Introduction}

One of the most exciting topics about nonperturbative aspects 
of superstring theory or 
M-theory is their various connections to gauge theories. 
Matrix theory, whose classical action is given by that of 
one-dimensional Yang-Mills theory with maximal supersymmetry, 
has been conjectured as a nonperturbative definition of M-theory 
by Banks, Fischler, Shenker and Susskind (BFSS)~\cite{BFSS}. 
Also, via its toroidal compactification on a circle 
and on a two-torus in the 
procedure of Taylor~\cite{Taylor}, it leads to a proposal 
for a nonperturbative definition 
of type IIA superstring theory~\cite{DVV,Banks-Seiberg} 
and type IIB superstring theory~\cite{Banks-Seiberg,Sethi-Susskind}, 
respectively. 
They are given by two- and three-dimensional maximally 
supersymmetric Yang-Mills (SYM) theories. 
In addition, another definition of type IIB superstring theory, 
which takes the form of the above supersymmetric Yang-Mills theory 
completely reduced to a point, 
has been proposed by Ishibashi, Kawai, Kitazawa and 
Tsuchiya (IKKT)~\cite{IKKT}. 

  However, in order to definitely extract 
nonperturbative aspects of string theory from 
these matrix string models, we need to know also about those 
of SYM theories. 
In general, it still remains to be an extremely difficult problem, 
so until now the analysis has been 
done in the limited cases. 
In energy sufficiently higher than the string scale, 
there exists the region such that 
physics can be well described by a perturbative expansion 
around the specific instanton configuration of the SYM theory 
which describes the world sheet geometry 
in the string scattering process~\cite{GHV}. 
In the case of the IIA matrix string theory, 
some interesting results coming from nonperturbative natures of 
strings have been discovered 
there --- existence of the minimal distance, 
limitation of the number of short strings and 
pair creation/annihilation of D-particles in intermediate states of the 
scattering process. 
Also, by mapping the SYM theories into cohomological field theories, 
the partition functions and some special operators, 
which must belong to the BRST-cohomology class 
in the cohomological theories, 
can be exactly calculated~\cite{MNS,KV,Sugino,KS}.  
There are some approaches to Matrix theory from the point of view of 
a generalized version of the conformal symmetry in four-dimensional 
SYM theory \cite{Yoneya}. 
Besides the analytical computations as above, 
the numerical analysis has been proceeded 
in IKKT model~\cite{Oda,Krauth,Krauth2,Ambjorn,Ambjorn2,Egawa,Bialas} 
and in BFSS Matrix theory~\cite{Janik}. 

  Recently, Matrix theory has been analyzed 
in the approach based on Gaussian approximation~\cite{KL,KLL}, 
where it is reported that the Gaussian approximation nicely captures 
the qualitative strong-coupling behavior 
in various simple models --- zero-dimensional 
$\phi^4$-theory, $\cN=2$  supersymmetric quantum mechanical systems --- 
and in Matrix theory case it gives 
the results fitting the predictions from 
the conjectured AdS/CFT duality \cite{Maldacena}. 

 Here, we concentrate into the bosonic Yang-Mills integrals 
in zero-dimension as a preparation to 
the analysis of the supersymmetric case (IKKT model), 
and calculate various quantities by using the Gaussian 
approximation and by its improved version. 
We calculate the free energy, the so-called space-time extent\footnote{In this paper, 
the name ``space-time extent'' is used for the quantity 
$\sqrt{\bra\frac{1}{N}\tr (X_{\mu})^2\ket}$.} and 
the expectation values of Polyakov loop and Wilson loop with 
a square-shaped contour.  
Our results match well with the results 
of either exact or numerical calculations reported before. 
Furthermore, by means of the improved mean field approximation, 
we calculate the expectation values of 
the Polyakov loop and the Wilson loop in the case of 
the length of the loop sufficiently large, 
which is guessed to reproduce the correct behaviors at least 
qualitatively.
Since in our knowledge the expectation values of the loops in this case 
have not ever been evaluated 
either analytically or numerically in literatures, 
the approximation scheme presented here might give some new insights. 
 
 The paper is organized as follows. 
In section 2, we explain two approximation schemes --- Gaussian 
approximation and improved mean field approximation --- by applying 
them to simple $\phi^4$-integral. 
We calculate the free energy and the correlator $\bra e^{iL\phi}\ket$. 
In the Gaussian approximation, it can be seen that 
the first few terms in the expansion around 
the Gaussian classical action yield a good approximation of the exact result. 
For the correlator $\bra e^{iL\phi}\ket$, this calculation can be trusted 
in the case of $L$ small, while, as $L$ increases, 
higher order terms becomes more and more relevant. 
So we have to consider another approximation scheme suitable 
for the case of $L$ large. 
The Gaussian approximation can be interpreted 
as a kind of mean field approximation. From the point of view, 
we can improve the mean field approximation 
to be appropriate for treating the large $L$ case. 
Then we find a precise correspondence between the solutions 
of the mean field equation and of the saddle point equation. 
As the consequence, it turns out that 
the improved mean field approximation gives the result which 
nicely captures the essential features of 
the asymptotic behavior evaluated by the saddle point method. 
In section 3, we review briefly some known results about the reduced Yang-Mills 
integrals for referring them later. Also, an exact result for the space-time 
extent in the $SU(2)$ case is added, which has not ever appeared in the literatures 
as far as we know. 
Section 4 is devoted to analysis of the reduced Yang-Mills integrals by means of 
the Gaussian approximation. Our results for the free energy and for the space-time 
extent fit well the exact results in the case of the dimensionality of space-time 
large. This is consistent with the fact that the Gaussian approximation is 
regarded as the mean field approximation. Furthermore, with respect to 
the Polyakov loop and the Wilson loop, our results match better with the numerical 
results given in ref. \cite{Ambjorn} in the region of the length of the loop 
smaller. In the case of general $N$, 
the formulas which we have obtained are relatively lengthy. 
However, after taking the 't Hooft like large $N$ limit,  
they become remarkably simplified. 
Looking at the several results obtained for various values of $N$ together, 
we can observe quite good scalings in particular in the region $N\geq 48$. 
In section 5, we continue the analysis using the improved mean field approximation. 
The result for the space-time extent shows a closer approximation to the exact result 
than that of the Gaussian approximation in the $SU(2)$ case. 
Also, the one-point functions of the Polyakov loop and the Wilson loop are obtained in 
the region of $L$ large. 
In our knowledge, they have not ever been evaluated either exactly or numerically 
in appropriate literatures. 
On the other hand, in the supersymmetric case, the Wilson loop amplitude 
for a few smaller values of $N$ 
has been numerically computed up to larger $L$ 
in ref. \cite{Krauth}\footnote{M. Staudacher informed us that 
the authors of ref. \cite{Krauth} had simulated the Wilson loop also in the 
bosonic Yang-Mills integrals, and that the similar result as in the supersymmetric 
case has been obtained \cite{Staudacher}. It is consistent to our result. 
We would like to thank M. Staudacher for his kindness.}. 
Here we find a strong resemblance between the result there and ours.  
We summarize the results obtained here 
and discuss about possible future directions in section 6.  
In appendix A, in order to confirm the correspondence between the solutions of 
the mean field equation and those of the saddle point equation, 
which is mentioned in section 2, we give another evidence which realizes the 
correspondence by investigating $\phi^6$-integral. 
Appendix B is devoted to a detailed derivation of the 't Hooft limit for 
the Polyakov loop and the Wilson loop.


\section{Gaussian and Mean Field Approximations}

  In this section, we explain two approximation schemes, 
which we will apply to Yang-Mills integrals later, 
by using a simple example ($\phi^4$-integral in zero-dimension). 
The first scheme is the so-called Gaussian approximation, 
which is discussed in the case 
of various supersymmetric quantum mechanical systems in ref.~\cite{KL,KLL}. 
The Gaussian approximation can be regarded as a kind of mean field approximation, 
and then we can consider some improvement for the calculation of 
various correlators. This is the second one. 
We will call it the improved mean field approximation.

\subsection{A Simple Example}

  We start with $\phi^4$-integral in zero-dimension with the action: 
$
S=\frac{1}{4g^2}\phi^4$.
The partition function $Z$ can be exactly calculated 
in terms of the Gamma-function: 
\beq
Z=\int^{\infty}_{-\infty}d\phi\; e^{-S}
=\sqrt{\frac{g}{2}}\;\Gamma\left(\frac14\right). 
\label{example-exact}
\eeq
Now, as a scheme which reproduces this result approximately, 
we consider the expansion around the Gaussian action 
$S_0=\frac{1}{2\sigma^2}\phi^2$: 
\beq
Z=\int^{\infty}_{-\infty}d\phi\; e^{-S_0}e^{-(S-S_0)}
=Z_0\bra e^{-(S-S_0)}\ket_0, 
\eeq
where $Z_0$ and $\bra \cdots \ket_0$ denote the partition function and 
the expectation value in the Gaussian theory. 
The width $\sigma$ is to be determined later. 
The free energy $F=-\ln Z$ is given by the form of the Cumulant expansion: 
\beq
F=F_0-\sum_{n=1}^{\infty}\frac{(-1)^n}{n!}\bra (S-S_0)^n\ket_{C,0}. 
\label{cumulantexpansion}
\eeq
The subscript $C$ means the connected expectation value 
in the Gaussian theory. 
The first few terms of the expansion are given by 
\bea
 & & F_0=-\ln Z_0=-\frac12\ln (2\pi\sigma^2), \nn \\
 & & \bra S-S_0\ket_0= \frac{3\sigma^4}{4g^2}-\frac12, \nn \\
 & & -\frac12\bra (S-S_0)^2\ket_{C,0}=
-3\frac{\sigma^8}{g^4}+\frac{3\sigma^4}{2g^2}-\frac14. 
\eea
Of course, the free energy $F$ is independent of $\sigma$ by definition. 
However, if we truncate the series (\ref{cumulantexpansion}) at some order, 
the value of the truncated series depends on $\sigma$. 
Generically summing up the whole series (\ref{cumulantexpansion}) seems to be 
a formidable task, thus it will be best to choose the value $\sigma$ so that 
the series exhibits a sufficiently fast convergence into the limit $F$ 
if it's possible 
and then to evaluate the truncated series at the optimized value $\sigma$.  

  Now let us determine $\sigma$ by the equation: 
\beq
\frac{\partial}{\partial\sigma^2}(F_0+\bra S-S_0\ket_0)=0. 
\label{sigma}
\eeq
It can be interpreted as the variational method, 
because the inequality 
$
F\leq F_0+\bra S-S_0\ket_0
$ 
holds and we consider the $\sigma$ minimizing 
the r.h.s. of the inequality\footnote{Eq. (\ref{sigma}) can also be 
regarded as 
mean field approximation as we will show in the next subsection.}. 
Eq. (\ref{sigma}) fixes $\sigma$ as $\sigma^2=\sqrt{g^2/3}$, 
then the free energy 
becomes 
\beq
F= -\frac12\ln g -\frac12\ln \frac{2\pi}{\sqrt{3}}-\frac14-\frac{1}{12} 
+O(\bra(S-S_0)^3\ket_{C,0}),
\label{example-freeenergy}
\eeq
where the third and fourth terms represent the contribution from 
$\bra S-S_0\ket_0$ and $-\frac12\bra (S-S_0)^2\ket_{C,0}$ respectively. 
Comparing the exact result 
\beq
F=-\frac12\ln g-\ln\frac{\Gamma(\frac14)}{\sqrt2}
=-\frac12\ln g-0.94144\cdots, 
\eeq
with the result up to the first order in $S-S_0$:``$-\frac12\ln g-0.89428\cdots$''  
and that up to the second order:``$-\frac12\ln g-0.97761\cdots$'', 
it seems that the series (\ref{example-freeenergy}) tends to converge 
into the exact value.

\paragraph{Correlator $\bra e^{iL\phi}\ket$}

    Next, we consider the expectation value of the operator $e^{iL\phi}$ 
(which corresponds to  Polyakov loop or  Wilson loop in reduced Yang-Mills 
integrals we will discuss later): 
\beq
\bra e^{iL\phi}\ket  = \frac1Z\int^{\infty}_{-\infty}d\phi\;e^{-S}e^{iL\phi}
 = \frac{\bra e^{-(S-S_0)}e^{iL\phi}\ket_0}{\bra e^{-(S-S_0)}\ket_0}.
\eeq
Expanding around the Gaussian theory, we have 
\beq
\bra e^{iL\phi}\ket = \bra e^{iL\phi}\ket_0 
+\sum_{n=1}^{\infty}\frac{(-1)^n}{n!}\bra(S-S_0)^n e^{iL\phi}\ket_{C,0}.
\eeq
At the value $\sigma^2=\sqrt{g^2/3}$ satisfying eq. (\ref{sigma}), 
the first few terms give 
\bea
\bra e^{iL\phi}\ket & = & e^{-\frac{L^2g}{2\sqrt3}} 
-\frac{1}{36}L^4g^2e^{-\frac{L^2g}{2\sqrt3}} \nn \\
 & & +\frac12\left(-\frac{2}{3\sqrt3}L^2g+\frac16L^4g^2
-\frac{1}{27\sqrt3}L^6g^3
+\frac{1}{1296}L^8g^4\right)e^{-\frac{L^2g}{2\sqrt3}}+\cdots, 
\label{example-loop}
\eea
where the second and third terms represent the contributions from the first 
and second order terms in the $(S-S_0)$-expansion. 
Note that $\bra (S-S_0)^ne^{iL\phi}\ket_{C,0}$ ($n>0$) has the form 
$p_n(L^2g) e^{-\frac{L^2g}{2\sqrt3}}$. $p_n(x)$ is a polynomial of $x$ 
of the degree $2n$ with no constant term. 
We can easily see this from a direct calculation except the point 
that the polynomial has no constant term. 
Including no constant term is understood from the fact that the connected 
correlator $\bra (S-S_0)^ne^{iL\phi}\ket_{C,0}$ vanishes as $L\limit 0$. 
Thus it can be expected that the expansion (\ref{example-loop}) 
gives a reasonable result at least qualitatively for the region of $L^2g$ small. 

  However, for the case of $L^2g$ large, since higher order terms 
become more and more 
dominant, we can not trust results obtained from the above expansion. 
In fact, in this case we can evaluate the asymptotic behavior 
of $\bra e^{iL\phi}\ket$ 
by using the saddle point method for the integral 
$\int^{\infty}_{-\infty}d\phi\;e^{-S}e^{iL\phi}$. 
The saddle point equation has three solutions: 
\beq
\phi_+=e^{\pi i/6}(Lg^2)^{1/3},\hspace{1cm} 
\phi_-=e^{5\pi i/6}(Lg^2)^{1/3}, \hspace{1cm}
\phi_0=e^{-\pi i/2}(Lg^2)^{1/3}. 
\label{saddle-point-solutions}
\eeq
We deform the integration contour so as to pass the two points $\phi_{\pm}$ 
along the steepest descent directions and 
evaluate the Gaussian integrals\footnote{For discussion in the next subsection, 
we note that the integrand 
at the saddle point $\phi_0$ is $e^{\frac34(L\sqrt{g})^{4/3}}$ 
which exhibits the behavior of blowing up as $L\limit\infty$.\label{saddle}}. 
The result after dividing by the partition function (\ref{example-exact}) is 
\beq
\bra e^{iL\phi}\ket\sim\frac{4}{\Gamma(\frac14)}\sqrt{\frac{\pi}{3}}
\frac{1}{(L\sqrt{g})^{1/3}}
e^{-\frac38(L\sqrt{g})^{4/3}}
\cos\left[\frac{3\sqrt3}{8}(L\sqrt{g})^{4/3}-\frac{\pi}{6}\right]. 
\label{example-saddlepoint2}
\eeq
The result (\ref{example-saddlepoint2}) exhibits 
qualitatively distinct behavior from the result obtained from 
the first few terms 
in the Gaussian approximation (\ref{example-loop}). 
In the next subsection, we discuss some improvement of the approximation 
which reproduces the behavior (\ref{example-saddlepoint2}) 
at least qualitatively.

\subsection{Improved Mean Field Approximation} 

  Bearing in mind the improvement, 
we begin with giving another interpretation 
of the Gaussian approximation as mean field approximation. 
Let us consider the mean field approximation to the $\phi^4$-integral 
by replacing $\phi^2$ in the $\phi^4$-action with the expectation value 
$\bra \phi^2 \ket$. The mean field action is 
\beq
S_M=\frac{1}{4g^2}\cdot 6\bra \phi^2 \ket \phi^2 +\frac{1}{\lambda} 
= \frac{1}{2\sigma^2}\phi^2 + \frac{1}{\lambda}, 
\eeq
where the factor ``6'' stands for the number of ways of contracting 
two $\phi$'s in the $\phi^4$-action, 
and the constant $\lambda$ was introduced 
for later convenience. 
The width of the Gaussian action $\sigma$ is related to 
the two-point function as 
\beq
\frac{1}{\sigma^2}=\frac{3}{g^2}\bra \phi^2 \ket. 
\label{example-width}
\eeq
The partition function is written as 
\beq
Z=\int^{\infty}_{-\infty}d\phi\; e^{-S_M}e^{-(S-S_M)}
=Z_M\bra e^{-(S-S_M)}\ket_M. 
\eeq
The partition function and the expectation values under the mean field theory 
with the classical action $S_M$ are denoted by $Z_M$ and $\bra \cdots \ket_M$. 
Since the factor $e^{-(S-S_M)}$ represents the difference 
between the original theory and the mean field theory, we want to take 
\beq
\bra e^{-(S-S_M)}\ket_M=1
\label{example-condition1}
\eeq
by choosing the parameters $\sigma$ and $\lambda$, 
so that the mean field theory 
realizes the original theory. 
Also, for the equivalence of both theories\footnote{Note that 
the equivalence stated here 
is according only to the partition function and the two point function, 
not to all correlators.} we require 
\beq
\bra \phi^2 \ket = \bra \phi^2 \ket_M. 
\label{example-condition2}
\eeq
The two conditions (\ref{example-condition1}) and (\ref{example-condition2}) 
determine the parameters $\sigma$ and $\lambda$. 
By combining with eq. (\ref{example-width}), 
the condition (\ref{example-condition2}) gives the same value of $\sigma$ as from 
the variational method (\ref{sigma}). 
Also, the condition (\ref{example-condition1}) means 
$
\sum_{n=1}^{\infty}\frac{(-1)^n}{n!}\bra (S-S_M)^n\ket_{C,M} =0. 
$
It is noted that $1/\lambda$ appears only in the first term of the l.h.s. of 
this equation because considering the connected correlators. From 
this equation, $\lambda$ is given by 
\beq
\frac{1}{\lambda}
= -\sum_{n=1}^{\infty}\frac{(-1)^n}{n!}\bra (S-S'_M)^n\ket_{C,M}
\label{example-lambda}
\eeq
where we put $S'_M=\frac{1}{2\sigma^2}\phi^2$. 
The free energy 
$
F =-\ln Z =-\ln Z_M = -\frac12 \ln (2\pi\sigma^2)+\frac{1}{\lambda} 
$
leads the identical result with the variational method applied to 
(\ref{cumulantexpansion}). 

  Next we consider some improvement of the above mean field approximation 
in the case of the unnormalized expectation value of the operator $\cO$. 
Let us repeat the same argument 
by regarding the unnormalized expectation value 
$
\bra \cO \ket'\equiv \int d\phi\;e^{-S}\cO 
$
as the partition function of a theory 
with $e^{-S}\cO$ being the Boltzmann weight. 
We take the following mean field action: 
\beq
S_M=\frac{1}{4g^2}\cdot 6\bra \phi^2 \ket^{\cO} \phi^2 +\frac{1}{\lt} 
= \frac{1}{2\st^2}\phi^2 + \frac{1}{\lt}. 
\eeq
Here, $\bra \cdots \ket^{\cO}$ stands for the expectation value 
under the Boltzmann weight $e^{-S}\cO$, i.e. for arbitrary operator $A$, 
\beq
\bra A \ket^{\cO} \equiv \frac{\int d\phi\;e^{-S}\cO A}{\int d\phi\;e^{-S}\cO} 
=\frac{\bra \cO A\ket}{\bra \cO \ket}. 
\eeq
As the result of the same argument as before, we obtain 
\beq
\frac{g^2}{3\st^2}=\frac{\bra \cO \phi^2\ket_M}{\bra \cO \ket_M},
\label{example2-condition-st}
\eeq
\beq
\frac{1}{\lt}=-\sum_{m=1}^{\infty}\frac{(-1)^m}{m!}\bra(S-S'_M)^m\ket_{C,M}
-\ln\left[1+\frac{1}{\bra\cO\ket_M}
\sum_{n=1}^{\infty}\frac{(-1)^n}{n!}\bra \cO (S-S'_M)^n\ket_{C,M}\right], 
\label{example2-lt}
\eeq
where $S'_M=\frac{1}{2\st^2}\phi^2$.

\paragraph{Correlator $\bra e^{iL\phi}\ket$}

  Let us apply this method to the case of $\cO=e^{iL\phi}$. From 
the condition (\ref{example2-condition-st}), 
\beq
x^3=-\frac13L^4g^2+x^2, 
\eeq
where $x\equiv L^2\st^2$. 
In the case of $L^2g$ large, this equation can be solved iteratively and 
the following three solutions are obtained: 
\bea 
x_{\pm} & = & 
e^{\pm\pi i/3}\left(\frac13L^4g^2\right)^{1/3}+\frac13+O((L\sqrt{g})^{-4/3}), \nn \\
x_0 & = & -\left(\frac13L^4g^2\right)^{1/3}+\frac13+O((L\sqrt{g})^{-4/3}). 
\label{example-solutions-x}
\eea
Corresponding to each solution of $x$, 
$\lt$ is determined by eq. (\ref{example2-lt}). 
In the form of the $(S-S'_M)$-expansion, it is expressed as 
\beq
\frac{1}{\lt}
=\frac{5}{12}x-\frac{1}{12}+O((L\sqrt{g})^{-4/3})+O((S-S'_M)^2)
\label{example-solution-lt1}
\eeq
up to the first order, and 
\beq
\frac{1}{\lt}
=\frac{23}{36}x+\frac{13}{36}+O((L\sqrt{g})^{-4/3})+O((S-S'_M)^3)
\label{example-solution-lt2}
\eeq
up to the second order. 
We will denote the three $\lt$'s 
corresponding to $x_{\pm}$ and $x_0$ by $\lt_{\pm}$ and $\lt_0$, respectively. 

 Here, we have three sets of the solutions $(\st,\lt)$. 
There is a problem --- which solution should we take? 
Also, does it make sense to adopt solutions more than one? 
If it does, how should we combine them? 
In this case, we can propose a prescription based on the following observation. 
The unnormalized correlation function is expressed in terms of $x$ and $\lt$ as 
\beq
\bra e^{iL\phi} \ket'=\frac1L\sqrt{2\pi x}\;\exp\left[-\frac12x-\frac{1}{\lt}\right]. 
\label{example2-loop}
\eeq
Examining the exponentiated factor in this equation, we are tempted to claim 
that these solutions reflect the structure of the solutions of the saddle point 
equation (\ref{saddle-point-solutions}). 
Actually, $x_0$ and $\lt_0$ lead to an unphysical amplitude 
which blows up as $\exp[c(L\sqrt{g})^{4/3}]$ with $c$ being a positive constant 
when $L\limit\infty$. This behavior is very close to that of the integrand at 
the saddle point $\phi_0$ (See footnote \ref{saddle} in the previous subsection.).   
Furthermore, we can see that both of $(x_+, \lt_+)$ and $\phi_-$ give the similar 
behavior as $\exp[-e^{\pi i/3}c(L\sqrt{g})^{4/3}]$ 
and that both of $(x_-, \lt_-)$ and $\phi_+$ lead 
$\exp[-e^{-\pi i/3}c(L\sqrt{g})^{4/3}]$. 
In particular, the solutions $(x, \lt)$ exactly reproduce the phase factors 
and the power of $L\sqrt{g}$ 
in the saddle point values of $-S+iL\phi$. From this fact, 
it will be plausible 
to assume that each solution $(x,\lt)$ certainly corresponds to 
each saddle point solution, 
although at present we have not found out the definite reason for the 
correspondence\footnote{In order to confirm this assumption further, 
we give another evidence for the case of $\phi^6$-integral in appendix A.}. 
If accepting this assumption, we can proceed further. 
As mentioned above, because the solution $(x_0,\lt_0)$ lead the unphysical solution, 
we discard it. 
Also, since the unnormalized expectation value 
$\bra e^{iL\phi} \ket'=\bra \cos (L\phi)\ket'$ is real, 
we combine the two solutions $(x_{\pm}, \lt_{\pm})$ with an same weight. 
Here, we determine the weight from the assumption.  
Let us take the combination same as what appears in the saddle point method. 
That is, we simply sum up the contributions from the two solutions with the weight one: 
\beq
\bra e^{iL\phi} \ket'=\frac1L\sqrt{2\pi 
x_+}\;\exp\left[-\frac12x_+-\frac{1}{\lt_+}\right]
+\frac1L\sqrt{2\pi x_-}\;\exp\left[-\frac12x_--\frac{1}{\lt_-}\right]. 
\label{example-prescription}
\eeq

  Finally, dividing by the partition function, we obtain the expression of the 
correlator. Then, for the precise cancellation of the vacuum graphs between 
the unnormalized correlator and the partition function, we need to use 
the result of the partition function up to the order same as that of 
the unnormalized correlator\footnote{Namely, it guarantees $\bra 1\ket=1$ 
in the result up to every order. 
This consideration becomes more relevant when treating the system 
with more degrees of freedom.}. 
Thus, for the result up to the first (second) order we use 
the result of the free energy up to the first (second) order. 
The final expression is 
\bea
\bra e^{iL\phi} \ket & = & 
2\cdot3^{1/12}e^{-u_n}\frac{1}{(L\sqrt{g})^{1/3}}
e^{-v_n\cdot3^{-1/3}(L\sqrt{g})^{4/3}}
\cos \left[v_n\cdot3^{1/6}(L\sqrt{g})^{4/3}-\frac{\pi}{6}\right]
\nn \\
 & & \times [1+O((L\sqrt{g})^{-4/3})+O((S-S'_M)^{n+1})], 
\eea
with up to the first order ($n=1$): $u_1=\frac{17}{36}$, $v_1=\frac{11}{24}$, 
and up to the second order ($n=2$): $u_2=\frac{29}{27}$, $v_2=\frac{41}{72}$. 

Now let us compare these with the result of the saddle point 
method (\ref{example-saddlepoint2}). 
First, the power behavior $1/(L\sqrt{g})^{1/3}$ and the power in the exponential 
and the cosine $(L\sqrt{g})^{4/3}$ just coincide. 
The coefficient of the exponential decay is 
\beq
\begin{array}{ll}
\frac{11}{24}\cdot3^{-1/3}=0.31779\cdots & \mbox{up to the 1st order} \\
\frac{41}{72}\cdot3^{-1/3}=0.39483\cdots & \mbox{up to the 2nd order} \\
\frac38=0.375                           & \mbox{saddle point method}, 
\end{array}
\eeq
where the second order result approaches closer to the saddle point result 
than the first order result. Also, the coefficient appearing in the argument 
of the cosine exhibits the similar behavior. 
On the other hand, the constant factor in front of the whole expression 
does not show a good result as long as 
looking at the first two orders: 
\beq
\begin{array}{ll}
2\cdot 3^{1/12}e^{-\frac{17}{36}}=1.3668\cdots & \mbox{up to the 1st order} \\
2\cdot 3^{1/12}e^{-\frac{29}{27}}=0.74873\cdots & \mbox{up to the 2nd order} \\
\frac{4}{\Gamma(\frac14)}\sqrt{\frac{\pi}{3}}=1.1289\cdots  
& \mbox{saddle point method}. 
\end{array}
\eeq
We need further examination of higher orders 
for convergence of the constant factor. 
The decay coefficient is determined by the first term in (\ref{example-solution-lt1}) 
(or (\ref{example-solution-lt2})) which is the leading in the case $L\sqrt{g}$ large. 
On the other hand, the constant factor is by the second term 
which is the subleading. From this point, we can understand that the decay coefficient 
converges faster than the constant factor. 
We can conclude that our scheme quite nicely reproduces the qualitative 
behavior of the saddle point result.


\section{Reduced Yang-Mills Integrals --- Exact Results}
\setcounter{equation}{0}

  In this section, we give some explanations about reduced Yang-Mills integrals, 
before applying the method discussed in the previous section. 
We review some known results as well as add an exact result for the space-time extent 
which have not been seen in appropriate references. 

  We consider the following bosonic Yang-Mills integral:
\beq
Z_{D,N}=\int \prod_{\mu=1}^{D}[dX_{\mu}]\; e^{-S(X)},
\label{partitionfunction-YM}
\eeq
with the Euclidean classical action:
$
S(X)=-\frac{1}{4g^2}\tr[X_{\mu},X_{\nu}][X_{\mu},X_{\nu}]$. 
The variables $X_{\mu}$'s are traceless hermitian matrices of the size $N\times N$. 
The indices $\mu$ and $\nu$ run from 1 to $D$. 
The normalization of the measure is determined by\footnote{This is nothing but 
the same normalization of the measure in ref. \cite{Krauth}. 
It can be easily seen by expanding $X$ by the basis $T^A$'s normalized as 
$\tr T^AT^B=\frac12\delta_{AB}$:
$
X=\sum_{A=1}^{N^2-1}X^AT^A$.}
$
\int [dX]\;e^{-\tr X^2}=1$. 
At first sight, this integral (\ref{partitionfunction-YM}) seems to lead 
an ill-defined result due to the integration over the infinite range 
along the flat directions. 
However, as shown in ref. \cite{Krauth,Krauth2}, 
in the $N=2$ case the integral can be performed exactly and 
it turns out to give the finite result when $D>4$:  
\beq
Z_{D,N=2}=\left\{\begin{array}{ll}
 \infty &(D\leq4) \\
 \frac12 g^{\frac32D}
\frac{\Gamma(\frac{D}{4})\Gamma(\frac{D}{4}-\frac12)\Gamma(\frac{D}{4}-1)}{\Gamma(
\frac{D}{2})\Gamma(\frac{D}{2}-\frac12)\Gamma(\frac{D}{2}-1)}
 & (D>4). 
\label{N=2exact1}
\end{array}\right.
\eeq
As is understood from intermediate steps in the integral, 
the finiteness is thanks to the suppression factor generated 
after integrating out the other variables 
than the variables along the flat directions.  
Furthermore, along the similar line we find the exact expression with respect to 
the square of the so-called space-time extent $\bra\frac12\tr(X_{\mu})^2\ket$:
\beq
\bra\frac12 \tr(X_{\mu})^2\ket=\frac{3g}{8}(D-2)
\frac{\Gamma(\frac{D}{4}-\frac32)}{\Gamma(\frac{D}{4}-1)}
\label{N=2exact2}
\eeq
for $D>6$. It is not well-defined for the case of $D\leq 6$. 
This result had not been obtained in literatures as far as we know. 

In the case of $N>2$, there are some insights from perturbative analysis. 
Let us start with considering the polar decomposition 
$X_{\mu}=V_{\mu}\Lambda_{\mu}V^{\dagger}_{\mu}$, where 
$\Lambda_{\mu}$'s are matrices of the eigenvalues of $X_{\mu}$'s: 
$\Lambda_{\mu}={\rm diag} (\lambda^1_{\mu},\cdots,\lambda^N_{\mu})$, 
and $V_{\mu}$'s are unitary matrices.  
The integrals of the unitary matrices are performed perturbatively 
by expanding around the unit matrix. 
As shown in ref. \cite{HNT}, from the formula of $Z_{D,N}$ 
after integrating $V_{\mu}$'s at the one-loop level, 
powercounting with respect to the $\Lambda_{\mu}$-integrals 
leads to the condition of the convergence 
of the integral for the large separation among the eigenvalues: 
\beq
N>\frac{D}{D-2} 
\label{convergence-condition}. 
\eeq
For the eigenvalue density of one of the $D$-matrices (say, $X_D$): 
$
\rho(\lambda)=\bra\frac1N\sum_{i=1}^N\delta(\lambda-\lambda^i_D)\ket$, 
a similar but more careful consideration about the $\Lambda_{\mu}$-integrals leads to 
the following asymptotic behavior: 
\beq
\rho(\lambda)\sim\lambda^{-2N(D-2)+3D-5}, 
\label{eigenvalue-density}
\eeq
which has been derived in ref. \cite{Krauth,Krauth2}. 
The formulas (\ref{convergence-condition}) and (\ref{eigenvalue-density}) are consistent 
with the results (\ref{N=2exact1}) and (\ref{N=2exact2}) in the $N=2$ case.

In the bosonic Yang-Mills integrals, we can also analyze 
by the $1/D$-expansion method 
as in ref. \cite{HNT}\footnote{The $1/D$-expansion method 
enables to evaluate various correlators and their large-$N$ scaling properties 
systematically as presented in ref. \cite{HNT}. 
At present, unfortunately it seems hard 
to systematically calculate highly 
complicated composite operators such as Polyakov loop and Wilson loop, which we discuss 
in this paper, by using the $1/D$-expansion formalism. 
The Gaussian and improved mean field approximations are not 
systematic from the point of view of the $1/D$-expansion, which is a weak point of 
these methods.
However, they enable to calculate those loop amplitudes 
and give a closed form at each order 
with respect to $(S-S_0)$-expansion and $(S-S'_M)$-expansion. 
As we will see, the results show good agreement with numerical results.}.
For a later discussion, we show the result for the square of the space-time extent: 
\beq
\bra \frac1N\tr(X_{\mu})^2\ket = \sqrt{\frac{DNg^2}{2}}\frac{N^2-1}{N^2}
\left[1+\frac1D\left(\frac76+\frac{1}{N^2-1}\right)+O(D^{-2})\right]. 
\label{1/D-expansion}
\eeq


\section{Reduced Yang-Mills Integrals --- Gaussian Approximation}
\setcounter{equation}{0}

  In this section, we apply the Gaussian approximation explained in section 2.1 
to the reduced Yang-Mills integrals. 
With respect to the partition function and to the space-time extent, 
our results nicely fit the known exact results of the $N=2$ case when $D$ is large. 
Also, we compute the one-point functions for the Polyakov loop and for the Wilson loop 
with a square-shaped contour. 
Our results turn out to reproduce well the numerical results 
in ref. \cite{Ambjorn} when the length of the loops 
is smaller.

\subsection{Partition Function and Space-Time Extent}

   First, we start with the partition function. Expanding around the Gaussian action: 
$
S_0=\frac{1}{\sigma^2}\tr (X_{\mu})^2$, 
we get the same expression as (\ref{cumulantexpansion})  
with 
\bea
 & & F_0=-\frac12(N^2-1)D\ln\sigma^2, \nn \\
 & & \bra S-S_0\ket_0=\frac{\sigma^4}{8g^2}N(N^2-1)D(D-1)-\frac12(N^2-1)D, \nn \\
 & & -\frac12\bra (S-S_0)^2\ket_{C,0}=-\frac{\sigma^8}{64g^4}N^2(N^2-1)D(D-1)(4D-1) \nn 
\\
 & &   \hspace{3.5cm}+\frac{\sigma^4}{4g^2}N(N^2-1)D(D-1)-\frac14(N^2-1)D. 
\eea
As the result of the variational method (\ref{sigma}), $\sigma$ is determined as 
\beq
\sigma^2=\sqrt{\frac{2g^2}{N(D-1)}}. 
\label{sigmaYM}
\eeq
As we saw in section 2, 
this approximation can be regarded as a kind of mean field approximation. 
So it is interesting to compare this with the exact result 
in the $N=2$ case (\ref{N=2exact1}) when $D$ is large. 
The exact result behaves as 
\beq
-\ln Z_{D,N=2}=\frac34D\ln D-\frac34D(\ln g^2+1)-2\ln 2 +O\left(\frac1D\right), 
\eeq
while the result by the Gaussian approximation becomes correspondingly 
\beq
F=\frac34D\ln D-\frac34D(\ln g^2+1)-\frac{21}{16}+O\left(\frac1D\right)
+O(\bra(S-S_0)^3\ket_{C,0}). 
\eeq
The first two terms in both formulas completely coincide. 
Also the $O(D^0)$-terms are $-2\ln 2=-1.3862\cdots$ and $-\frac{21}{16}=-1.3125\cdots$. 
It seems that the Gaussian approximation quite nicely reproduces the exact result 
in the large $D$ case. Furthermore, we can compare with the numerical result for 
smaller $D$ and $N$'s (Table 1 in the second paper of ref. \cite{Krauth2}), from 
which the values of $\tilde{F}_{(D,N)}\equiv -\ln Z+\frac14D(N^2-1)\ln(2g^2)$ are read 
off as 
\beq
\begin{array}{llll}
\tilde{F}_{(3,4)}=4.98\cdots, & \tilde{F}_{(3,5)}=13.8\cdots, & 
\tilde{F}_{(3,6)}=26.1\cdots, &  \\  
\tilde{F}_{(4,3)}=6.27\cdots, & \tilde{F}_{(4,4)}=17.4\cdots, & 
\tilde{F}_{(4,5)}=33.8\cdots, & \tilde{F}_{(4,6)}=56.2\cdots. 
\end{array}
\eeq
Let us look the quantity 
$\Delta_{(D,N)}\equiv (\tilde{F}'_{(D,N)}-\tilde{F}_{(D,N)})/\tilde{F}_{(D,N)}$ 
where $\tilde{F}'_{(D,N)}$ is the quantity in the Gaussian approximation 
corresponding to $\tilde{F}_{(D,N)}$. Then, 
\beq
\begin{array}{llll}
\Delta_{(3,4)}=0.59\cdots, & \Delta_{(3,5)}=0.21\cdots, & \Delta_{(3,6)}=0.11\cdots, \\
\Delta_{(4,3)}=0.21\cdots, & \Delta_{(4,4)}=0.063\cdots, & \Delta_{(4,5)}=0.036\cdots, &
\Delta_{(4,6)}=0.021\cdots. 
\end{array}
\eeq
This shows a tendency of better agreement not only for larger $D$ 
but also for larger $N$. 
   
   Next, let us examine the square of 
the space-time extent $\bra\frac1N\tr(X_{\mu})^2\ket$. We have 
\beq
\bra\frac1N\tr(X_{\mu})^2\ket=\bra\frac1N\tr(X_{\mu})^2\ket_0
+\sum_{n=1}^{\infty}\frac{(-1)^n}{n!}\bra (S-S_0)^n\frac1N\tr(X_{\mu})^2\ket_{C,0}, 
\eeq
with 
\bea
 & & \bra\frac1N\tr(X_{\mu})^2\ket_0=\frac{\sigma^2}{2}\frac{N^2-1}{N}D, \nn \\
 & & -\bra (S-S_0)\frac1N\tr(X_{\mu})^2\ket_{C,0}=
-\frac{\sigma^6}{4g^2}(N^2-1)D(D-1)+\frac{\sigma^2}{2}\frac{N^2-1}{N}D, \nn \\
 & & \frac12\bra (S-S_0)^2\frac1N\tr(X_{\mu})^2\ket_{C,0}=
\frac{\sigma^{10}}{16g^4}N(N^2-1)D(D-1)(4D-1) \nn \\ 
 & & \hspace{5cm} 
-\frac{3\sigma^6}{4g^2}(N^2-1)D(D-1)+\frac{\sigma^2}{2}\frac{N^2-1}{N}D. 
\eea
At the value (\ref{sigmaYM}), the first order term vanishes, and we obtain 
the expression 
\beq
\bra\frac1N\tr(X_{\mu})^2\ket=\sqrt{\frac{Ng^2}{2(D-1)}}\frac{N^2-1}{N^2}D
\left[1+\frac{3}{2(D-1)}+O((S-S_0)^3)\right]. 
\eeq
When $D$ is large, the second order term is suppressed by the factor $\frac{1}{D-1}$ 
comparing to the zeroth order term. 
It is consistent to the picture of the Gaussian approximation 
as the mean field approximation. 
In the $N=2$ case, this has the following large $D$ expansion: 
\beq
\bra\frac12\tr(X_{\mu})^2\ket=\frac{3g}{4}\sqrt{D}\left[1+\frac{2}{D}+O(D^{-2})
+O((S-S_0)^3)\right]. 
\label{GaussianlargeD}
\eeq
On the other hand, the exact result (\ref{N=2exact2}) does 
\beq
\bra\frac12\tr(X_{\mu})^2\ket=\frac{3g}{4}\sqrt{D}\left[1+\frac{3}{2D}+O(D^{-2})\right]. 
\label{N=2exactlargeD}
\eeq
We find some difference between the $O(D^{-1/2})$-terms. 
As will be seen in the next section, 
the situation becomes better when applying the improved mean field approximation.

\subsection{Polyakov Loop}

  Here we consider the expectation value of the operator of a loop of the length $L$ 
winding in one direction (say, the first direction)
$\hat{P}(L)=\frac1N\tr e^{iLX_1}$, which we will call Polyakov loop. 
The expectation value is expanded as 
\beq
\bra\hat{P}(L)\ket=\bra\hat{P}(L)\ket_0
+\sum_{n=1}^{\infty}\frac{(-1)^n}{n!}\bra (S-S_0)^n\hat{P}(L)\ket_{C,0}. 
\label{expansionPolyakov}
\eeq
For calculating each term in the r.h.s. 
it convenient to consider the Gaussian integral 
over the hermitian matrices (including the trace part) 
and to use the orthogonal polynomial method for 
the integrals with respect to the eigenvalues. 
Let us introduce the hermitian matrix $Y$ by adding the trace part to 
the traceless hermitian matrix $X$: 
$
Y=X+y\frac{1}{\sqrt{2N}}{\bf 1}_N$, 
where $y\in {\bf R}$. 
Also, the measure $[dY]$ is normalized by 
$
\int [dY]\; e^{-\tr Y^2}=1$. 
We consider the following expectation value in the hermitian Gaussian integral: 
\beq
\bra\frac1N\tr e^{iLY}\ket_0\equiv 
\frac{\int[dY]\;e^{-\frac{1}{\sigma^2}\tr Y^2}\frac1N\tr e^{iLY}}{\int[dY]\;
e^{-\frac{1}{\sigma^2}\tr Y^2}}
\label{Polyakov00}. 
\eeq
The Gaussian weight can be factorized into the product of the trace part 
and the traceless part. After integrating out the trace part, eq. (\ref{Polyakov00}) 
turns out to be related to the Gaussian expectation value of the Polyakov loop as 
\beq
\bra\frac1N\tr e^{iLY}\ket_0=e^{-\frac{L^2\sigma^2}{4N}}\bra\hat{P}(L)\ket_0. 
\label{Polyakov01}
\eeq

 Also, the l.h.s. of this equation reduces to the integrals 
with respect to the eigenvalues $\{\lambda_i\}$ of $Y$, and it can be easily evaluated 
by using the orthogonal polynomial method \cite{IZ}\footnote{Unfortunately, 
in the standard framework of the $1/D$-expansion in ref. \cite{HNT}, 
it seems hard to apply the orthogonal polynomial method, 
which is the difficulty pointed out before.}.  
It is translated into the language in the quantum mechanical system 
of a harmonic oscillator. 
Connected correlators among the $U(N)$-invariant operators $\tr F_k(Y)$ 
($k=1,2,\cdots$) are expressed as 
\bea
 & & \bra \tr F_1(Y)\ket_0=\sum_{n=0}^{N-1}\bra n|F_1(\hat{\lambda})|n\ket, \nn \\ 
 & & \bra \tr F_1(Y)\;\tr F_2(Y)\ket_{C,0}=
\Tr[\Pi_NF_1(\hat{\lambda})(1-\Pi_N)F_2(\hat{\lambda})], \nn \\
 & & \bra \tr F_1(Y)\;\tr F_2(Y)\;\tr F_3(Y)\ket_{C,0}=
\Tr [\Pi_NF_1(\hat{\lambda})F_2(\hat{\lambda})F_3(\hat{\lambda})
-\Pi_NF_1(\hat{\lambda})\Pi_NF_2(\hat{\lambda})F_3(\hat{\lambda}) \nn \\
 & & \hspace{3cm}-F_1(\hat{\lambda})\Pi_NF_2(\hat{\lambda})\Pi_NF_3(\hat{\lambda})
-\Pi_NF_1(\hat{\lambda})F_2(\hat{\lambda})\Pi_NF_3(\hat{\lambda}) \nn \\ 
 & & \hspace{3cm}+\Pi_NF_1(\hat{\lambda})\Pi_NF_2(\hat{\lambda})\Pi_NF_3(\hat{\lambda})
+\Pi_NF_1(\hat{\lambda})\Pi_NF_3(\hat{\lambda})\Pi_NF_2(\hat{\lambda})], \nn \\
 & &\hspace{2cm} \cdots, 
\label{connectedcorrelators}
\eea
where the creation and annihilation operators 
$\hat{a}$ and $\hat{a}^{\dagger}$ appearing in  $\hat{\lambda}$ 
as $\hat{\lambda}=\frac{\sigma}{\sqrt{2}}(\hat{a}+\hat{a}^{\dagger})$ satisfy 
$[\hat{a}, \hat{a}^{\dagger}]=1$. 
The states $\{|n\ket\}_{n=0,1,\cdots}$ form an orthonormal basis in the Fock space 
of the system of the quantum harmonic oscillator: 
$
\hat{a}|n\ket=|n-1\ket \sqrt{n}$, 
$\hat{a}^{\dagger}|n\ket=|n+1\ket \sqrt{n+1}$. 
``$\Tr$'' means the trace operation over the infinite dimensional Fock space. 
Also, ``$\Pi_N$'' stands for the projection operator into the $N$-dimensional space: 
$\Pi_N=\sum_{n=0}^{N-1}|n\ket\bra n|$.
By using the first formula in eqs. (\ref{connectedcorrelators}) we find 
\beq
\bra\frac1N\tr e^{iLY}\ket_0
=e^{-\frac{L^2\sigma^2}{4}}F\left(1-N,2;\frac{L^2\sigma^2}{2}\right), 
\label{Polyakov02}
\eeq
where $F$ is the confluent hypergeometric function: 
$$
F(\alpha,\beta;z)=\sum_{n=0}^{\infty}\frac{\alpha(\alpha+1)\cdots (\alpha+n-1)}{
\beta(\beta+1)\cdots(\beta+n-1)}\frac{z^n}{n!}. 
$$
Thus from eqs. (\ref{Polyakov01}) and (\ref{Polyakov02}), we obtain 
\beq
\bra\hat{P}(L)\ket_0=e^{-\frac{L^2\sigma^2}{4}(1-\frac1N)}
F\left(1-N,2;\frac{L^2\sigma^2}{2}\right).
\label{0thPolyakov} 
\eeq

\paragraph{First Order Term} 

  Next, we compute the first order term, namely $-\bra(S-S_0)\hat{P}(L)\ket_{C,0}$. 
Let us consider $\bra S_0 \hat{P}(L)\ket_0$: 
$
\bra S_0 \hat{P}(L)\ket_0=\frac{D-1}{D}\bra S_0\ket_0\bra\hat{P}(L)\ket_0+
\bra\frac{1}{\sigma^2}\tr (X_1)^2\hat{P}(L)\ket_0$. 
We pass to the $Y$-integral in order to evaluate $\bra\tr (X_1)^2\hat{P}(L)\ket_0$. 
Then, 
\bea
\bra S_0 \hat{P}(L)\ket_0 & = & 
e^{\frac{L^2\sigma^2}{4N}}\bra\frac{1}{\sigma^2}\tr Y^2\frac1N\tr e^{iLY}\ket_{C,0}
+\bra\frac{1}{\sigma^2}\tr Y^2\ket_0\bra\hat{P}(L)\ket_0 \nn \\
 & & +\frac{D-1}{D}\bra S_0\ket_0\bra\hat{P}(L)\ket_0
-\left(\frac12-\frac{L^2\sigma^2}{4N}\right)\bra\hat{P}(L)\ket_0.
\eea
By making use of the second formula in eqs. (\ref{connectedcorrelators}), 
the connected correlator in the first term can be calculated. 
Evaluating $\bra S \hat{P}(L)\ket_0$ similarly, we eventually obtain 
\bea
\lefteqn{-\bra(S-S_0)\hat{P}(L)\ket_{C,0}=
e^{-\frac{L^2\sigma^2}{4}(1-\frac1N)}\left(\frac{\sigma^4}{2g^2}N(D-1)-1\right)
\frac{L^2\sigma^2}{8}}\nn \\
 & & \;\times\left[(N-1)F\left(2-N,3;\frac{L^2\sigma^2}{2}\right)
+(N+1)F\left(1-N,3;\frac{L^2\sigma^2}{2}\right)\right]\nn \\
 & & -\left(\frac{\sigma^4}{2g^2}N(D-1)-1\right)
\frac{L^2\sigma^2}{4N}\bra\hat{P}(L)\ket_0. 
\label{1stPolyakov}
\eea

\paragraph{Second Order Term} 

  For the second order term $\frac12\bra(S-S_0)^2\hat{P}(L)\ket_{C,0}$, 
we can also compute in the similar manner as in the first order case. 
After a relatively long but straightforward calculation, 
we arrive at the following result: 
\bea
\lefteqn{\frac12\bra(S-S_0)^2\hat{P}(L)\ket_{C,0}=}\nn \\
 & & e^{-\frac{L^2\sigma^2}{4}(1-\frac1N)}\left[1-\frac{\sigma^4}{g^2}N(D-1)
-\frac{\sigma^8}{4g^4}(N-3)(D-1)+\frac{\sigma^8}{4g^4}N^2(D-1)^2\right]
\frac{L^4\sigma^4}{32N} \nn \\
 & & \times\left[-\left(\begin{array}{c} N \\ 4\end{array}\right)
F\left(4-N,5;\frac{L^2\sigma^2}{2}\right)
-\left(\begin{array}{c} N+1 \\ 4\end{array}\right)
F\left(3-N,5;\frac{L^2\sigma^2}{2}\right)\right] \nn \\
 &+& e^{-\frac{L^2\sigma^2}{4}(1-\frac1N)}\left[1-\frac{\sigma^4}{g^2}N(D-1)
+\frac{\sigma^8}{4g^4}(N+3)(D-1)+\frac{\sigma^8}{4g^4}N^2(D-1)^2\right]
\frac{L^4\sigma^4}{32N} \nn \\
 & & \times\left[\left(\begin{array}{c} N+2 \\ 4\end{array}\right)
F\left(2-N,5;\frac{L^2\sigma^2}{2}\right)
+\left(\begin{array}{c} N+3 \\ 4\end{array}\right)
F\left(1-N,5;\frac{L^2\sigma^2}{2}\right)\right] \nn \\
 &+& e^{-\frac{L^2\sigma^2}{4}(1-\frac1N)}
\frac{\sigma^8}{8g^4}(D-1)\frac{L^4\sigma^4}{4N}
\left[-\left(\begin{array}{c} N\\3\end{array}\right)
F\left(3-N,4;\frac{L^2\sigma^2}{2}\right)\right. \nn \\
 & & \left. -\left(\begin{array}{c} N+1\\3\end{array}\right)
F\left(2-N,4;\frac{L^2\sigma^2}{2}\right)
-\left(\begin{array}{c} N+2\\3\end{array}\right)
F\left(1-N,4;\frac{L^2\sigma^2}{2}\right) \right] \nn \\
 &-& e^{-\frac{L^2\sigma^2}{4}(1-\frac1N)}\left[\left\{1
-\frac{3\sigma^4}{2g^2}N(D-1)+\frac{\sigma^8}{8g^4}N(2N-1)(D-1)
+\frac{\sigma^8}{8g^4}N^2(D-1)(4D-3)\right\}\right.\nn \\
 & & \left.\times\frac{L^2\sigma^2}{4N}
+\left\{1-\frac{\sigma^4}{2g^2}N(D-1)\right\}^2
\frac{L^4\sigma^4}{16N^2}\right]\left(\begin{array}{c} N\\2\end{array}\right)
F\left(2-N,3;\frac{L^2\sigma^2}{2}\right) \nn \\
 &-& e^{-\frac{L^2\sigma^2}{4}(1-\frac1N)}\left[\left\{1
-\frac{3\sigma^4}{2g^2}N(D-1)+\frac{\sigma^8}{8g^4}N(2N+1)(D-1)
+\frac{\sigma^8}{8g^4}N^2(D-1)(4D-3)\right\}\right.\nn \\
 & & \left.\times\frac{L^2\sigma^2}{4N}
+\left\{1-\frac{\sigma^4}{2g^2}N(D-1)\right\}^2
\frac{L^4\sigma^4}{16N^2}\right]\left(\begin{array}{c} N+1\\2\end{array}\right)
F\left(1-N,3;\frac{L^2\sigma^2}{2}\right) \nn \\
 &+& \left[\left\{1-\frac{3\sigma^4}{2g^2}N(D-1)+\frac{\sigma^8}{2g^4}N^2D(D-1)\right\}
\frac{L^2\sigma^2}{4N}+\left\{1-\frac{\sigma^4}{2g^2}N(D-1)\right\}^2
\frac{L^4\sigma^4}{32N^2}\right] \nn \\
 & & \times \bra\hat{P}(L)\ket_0. 
\label{2ndPolyakov}
\eea

\paragraph{Gaussian Approximation}

   Let us evaluate the first three terms in the expansion (\ref{expansionPolyakov}), 
i.e. eqs. (\ref{0thPolyakov}), (\ref{1stPolyakov}) and (\ref{2ndPolyakov}), 
at the value (\ref{sigmaYM}). 
The first order term (\ref{1stPolyakov}) vanishes, 
and the second order term (\ref{2ndPolyakov}) takes the form 
\bea
\lefteqn{\frac12\bra(S-S_0)^2\hat{P}(L)\ket_{C,0}=
e^{-\frac{L^2\sigma^2}{4}(1-\frac1N)}}\nn \\
 &\times &\left\{ \frac{N-3}{N^2(D-1)}
\frac{L^4\sigma^4}{32N} 
\left[\left(\begin{array}{c} N \\ 4\end{array}\right)
F\left(4-N,5;\frac{L^2\sigma^2}{2}\right)
+\left(\begin{array}{c} N+1 \\ 4\end{array}\right)
F\left(3-N,5;\frac{L^2\sigma^2}{2}\right)\right] \right.\nn \\
 & & +\frac{N+3}{N^2(D-1)}
\frac{L^4\sigma^4}{32N}
\left[\left(\begin{array}{c} N+2 \\ 4\end{array}\right)
F\left(2-N,5;\frac{L^2\sigma^2}{2}\right)
+\left(\begin{array}{c} N+3 \\ 4\end{array}\right)
F\left(1-N,5;\frac{L^2\sigma^2}{2}\right)\right] \nn \\
 & & +\frac{1}{2N^2(D-1)}
\frac{L^4\sigma^4}{4N}
\left[-\left(\begin{array}{c} N\\3\end{array}\right)
F\left(3-N,4;\frac{L^2\sigma^2}{2}\right)\right. \nn \\
 & & \left. -\left(\begin{array}{c} N+1\\3\end{array}\right)
F\left(2-N,4;\frac{L^2\sigma^2}{2}\right)
-\left(\begin{array}{c} N+2\\3\end{array}\right)
F\left(1-N,4;\frac{L^2\sigma^2}{2}\right) \right] \nn \\
 & & -\frac{3N-1}{2N(D-1)}
\frac{L^2\sigma^2}{4N}
\left(\begin{array}{c} N\\2\end{array}\right)
F\left(2-N,3;\frac{L^2\sigma^2}{2}\right) \nn \\
 & & \left.-\frac{3N+1}{2N(D-1)}
\frac{L^2\sigma^2}{4N}
\left(\begin{array}{c} N+1\\2\end{array}\right)
F\left(1-N,3;\frac{L^2\sigma^2}{2}\right)\right\} \nn \\
 &+& \frac{2}{D-1}\frac{L^2\sigma^2}{4N}\bra\hat{P}(L)\ket_0, 
\label{2ndPolyakov2}
\eea
where $\sigma^2=\sqrt{\frac{2g^2}{N(D-1)}}$. 
We can observe that the factor $\frac{1}{D-1}$ exists 
in every term in the second order result. 
Again, it is consistent with our picture as the mean field approximation. 
  
  In the case of $D=4$, there are some numerical results about the Polyakov loop 
and the Wilson loop reported in ref. \cite{Ambjorn}. 
So we can compare our result with the numerical one. 
Let us take $Ng^2=48$ and consider the quantity for various values of $N$ 
keeping $Ng^2$ fixed, which gives the same setting as in ref. \cite{Ambjorn}. 
Figure 13 in ref. \cite{Ambjorn} is the result to be compared with ours. 
The variable $k/\sqrt{g}$ appearing there corresponds to $L$ in our setting. 
In Fig. 1 we show the result of the zeroth order alone 
and that summed up to the second order 
for the Polyakov loop when $N=16$. 
It can be seen that our result up to the second order 
nicely reproduces the numerical result 
for the region of smaller $L$ (up to about 1.0). There are some 
differences between them in larger $L$, 
where in our analysis higher order terms are considered to 
become more and more important. (See also Fig. 2.)  

\begin{figure}
\epsfxsize=9cm \epsfysize=6cm
\centerline{\epsfbox{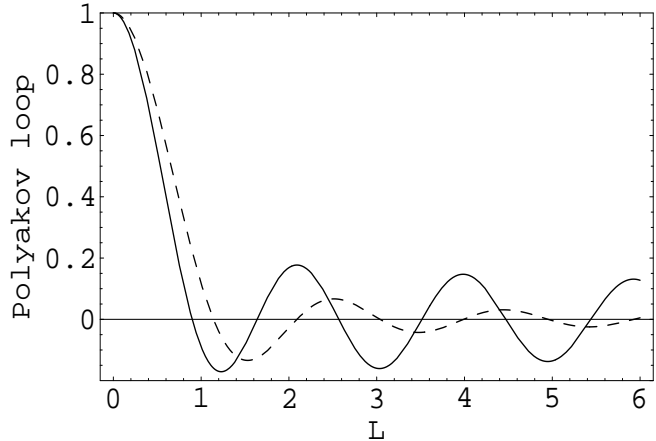}}
\caption{Polyakov loop in the case $D=4$ and $N=16$. 
The dotted line shows the contribution from the zeroth order alone, and 
the solid line shows that up to the second order.}
\end{figure}
  Furthermore, the result in ref. \cite{Ambjorn} exhibits a really good scaling behavior 
against various $N$'s. It can be seen also in our result. 
First let us consider the limit $N\limit\infty$ with $Ng^2$ fixed. 
In this limit, it turns out that 
the confluent geometric functions reduce to the Bessel functions 
and that the formula becomes considerably simplified as\footnote{We give some 
detailed explanation with respect to the derivation of this limit in appendix B.}.  
\beq
\bra\hat{P}(L)\ket=\frac{1}{\sqrt{\alpha}}J_1(2\sqrt{\alpha})
+\frac{1}{2(D-1)}\left[J_4(2\sqrt{\alpha})-3J_2(2\sqrt{\alpha})\right]+O((S-S_0)^3), 
\label{PolyakovlargeN}
\eeq
where $\alpha=\frac{L^2G}{\sqrt{2(D-1)}}$, and $G$ stands for the fixed value 
$\sqrt{N}g$. 
The first and the second terms in this formula come from the contributions of 
the zeroth and the second order terms, respectively. 
In $D=4$, we plot this result with $G^2=48$ in Fig. 2. 
Comparing it with the curve in Fig. 1, it seems to suggest a good scaling 
behavior. In fact, we plot our results for various values 
of $N$ in Fig. 3 ($N=16$, 48, 100, 400 and $\infty$).  
In Fig. 3 we can see some slight differences between the curve of $N=16$ 
and the others for $L>3$. Among the curves for $N\ge 48$, however the scaling is 
too good to observe any difference for various $N$ in the region 
$L<6$. 

\begin{figure}
\epsfxsize=9cm \epsfysize=6cm
\centerline{\epsfbox{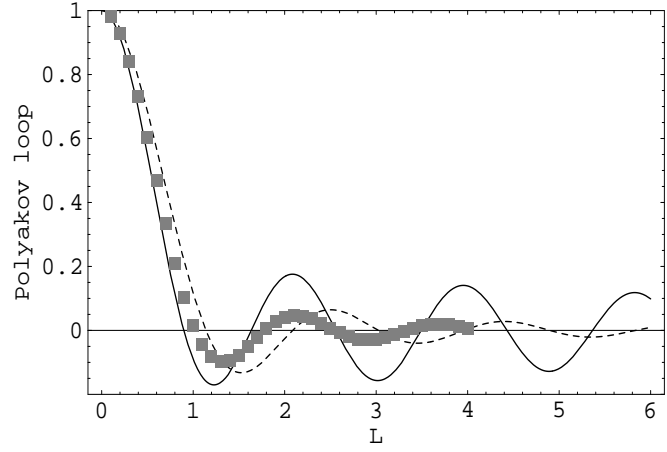}}
\caption{Polyakov loop in the case $D=4$ and $N=\infty$. 
The dotted line shows the contribution from the zeroth order alone, and 
the solid line shows that up to the second order.
The gray boxes indicate some points of the numerical data 
for the $N=48$ case in Figure 13 in ref. \cite{Ambjorn}.}
\end{figure}
\begin{figure}
\epsfxsize=9cm \epsfysize=6cm
\centerline{\epsfbox{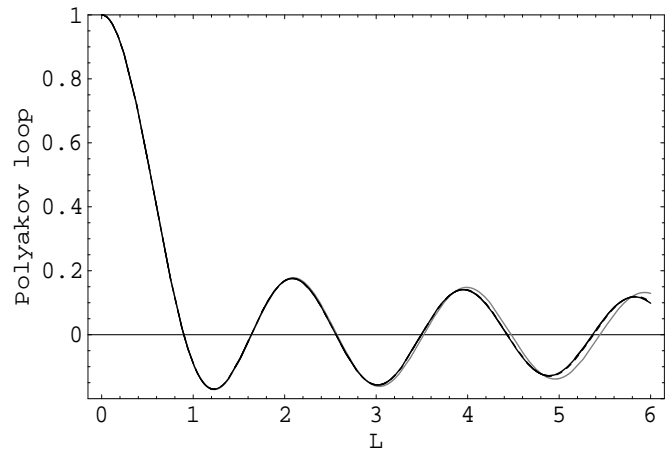}}
\caption{Polyakov loops up to the second order 
for $(D,N)=(4,16)$, $(4,48)$, $(4,100)$, $(4,400)$ and $(4,\infty)$. 
The gray line shows the case $(4,16)$, the dotted line $(4,48)$, 
the dashed line $(4,100)$, 
the dot-dashed line $(4,400)$, and the solid line $(4,\infty)$.}
\end{figure}

   Since we have obtained the concrete formula for general $D$ and $N$, 
investigating the behavior for other $D$'s can be immediately done. 
For example, we plot the $D=10$ case and the $D=26$ case for various $N$'s 
in Fig. 4. 
Both casess still exhibit quite good scaling behaviors. 
As $D$ increases, the curve tends to broaden and to make the amplitude of the 
oscillation decreased. 
We can understand the tendency by seeing the large $N$ result (\ref{PolyakovlargeN}). 
The former can be understood from the fact that $L$ is contained only 
through the combination of $\alpha$, 
and the latter from the suppression of the second term when $D$ being large. 

\begin{figure}
\epsfxsize=9cm \epsfysize=6cm
\centerline{\epsfbox{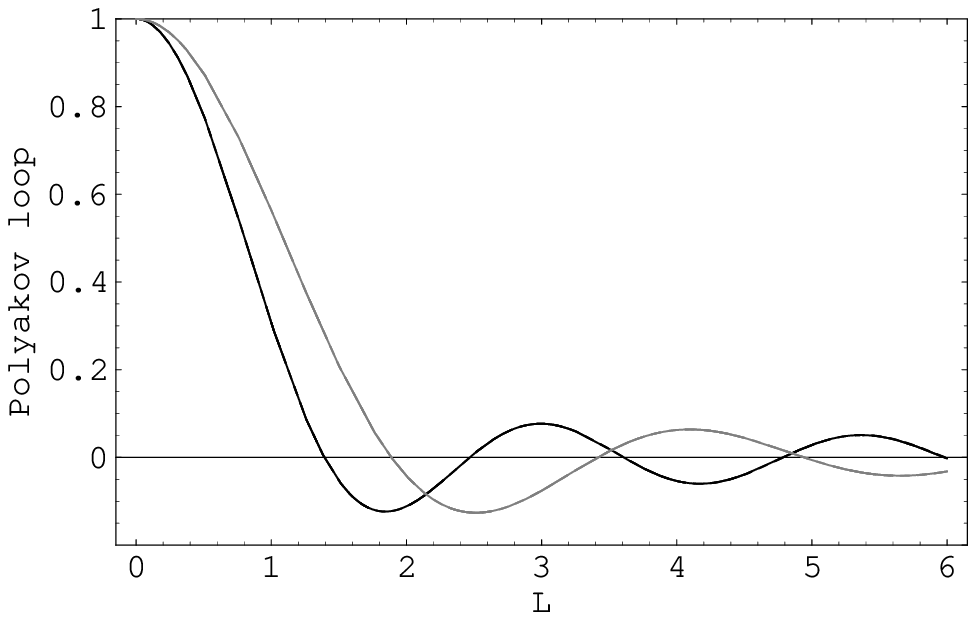}}
\caption{Polyakov loops up to the second order 
for $(D,N)=(10,48)$, $(10,100)$, $(10,400)$, $(10,\infty)$, $(26,48)$, $(26,100)$, 
$(26,400)$ and $(26,\infty)$. 
The dotted line shows the case $(10,48)$, the dashed line $(10,100)$, 
the dot-dashed line $(10,400)$, the solid line $(10,\infty)$, 
the gray dotted line $(26,48)$, the gray dashed line $(26,100)$, 
the gray dot-dashed line $(26,400)$, and the gray solid line $(26,\infty)$.}
\end{figure}

  Thus we have seen that the Gaussian approximation reproduces well 
the numerical result of the Polyakov loop in smaller $L$ region. 
How about the Polyakov loop in larger $L$ region? From the same reason as 
the case of the simple example in section 2, the Gaussian approximation can not 
be trusted. 
In section 5, we will consider it by using the improved mean field approximation, 
and derive the asymptotic behavior of the Polyakov loop approximately.

\subsection{Wilson Loop}

   We consider the expectation value of the rectangular Wilson loop of 
the size $L\times L$: $\hat{W}(L)=\frac1N\tr(e^{iLX_1}e^{iLX_2}e^{-iLX_1}e^{-iLX_2})$. 
Here we evaluate the first two terms in the expansion: 
\beq
\bra\hat{W}(L)\ket=\bra\hat{W}(L)\ket_0
+\sum_{n=1}^{\infty}\frac{(-1)^n}{n!}\bra (S-S_0)^n\hat{W}(L)\ket_{C,0}. 
\label{expansionWilson}
\eeq

\paragraph{Leading Order Term}

  First let us evaluate the leading (zeroth) order term $\bra\hat{W}(L)\ket_0$, 
which is written as 
\beq
\bra\hat{W}(L)\ket_0=\frac1NI(L)_{ijkl}I(L)_{jkli}, 
\eeq
with $I(L)_{ijkl}\equiv\bra(e^{iLX_1})_{ij}(e^{-iLX_1})_{kl}\ket_0$. From the 
invariance 
of the measure $[dX_1]$ under $X_1\limit UX_1U^{\dagger}$ with $U$ being an arbitrary 
$SU(N)$ matrix, the quantity  $I(L)_{ijkl}$ must satisfy 
\beq
I(L)_{ijkl}=U_{ii'}(U^{\dagger})_{j'j}U_{kk'}(U^{\dagger})_{l'l}I(L)_{i'j'k'l'}. 
\eeq
It determines the structure of the indices of $I(L)_{ijkl}$ as\footnote{In the $N=2$ 
case, 
the tensor $\epsilon_{ij}$ seems to need to be taken into account. 
However, we do not have to worry about it because of the identity:
$\epsilon_{ik}\epsilon_{jl}=\delta_{ij}\delta_{kl}-\delta_{il}\delta_{jk}$.}
\beq
I(L)_{ijkl}=A(L)\delta_{ij}\delta_{kl}+B(L)\delta_{il}\delta_{jk}, 
\eeq
where $A(L)$ and $B(L)$ are $SU(N)$-invariant. Noting $I(L)_{ijjl}=\delta_{il}$ 
and $I(L)_{ijki}=\delta_{jk}$, $A(L)$ and $B(L)$ are determined. Thus we can rewrite 
$\bra\hat{W}(L)\ket_0$ in terms of correlators in the Gaussian one-matrix model: 
\beq
\bra\hat{W}(L)\ket_0=1-\frac{N^2}{N^2-1}
\left[\bra\frac1N\tr e^{iLX_1}\frac1N\tr e^{-iLX_1}\ket_0-1\right]^2.
\eeq
Considering the integral over the hermitian matrix $Y$ as before, the above correlator 
can be evaluated, and we obtain the expression 
\beq
\bra\hat{W}(L)\ket_0=1-\frac{N^2}{N^2-1}\left[-1+\frac1N+\frac{2}{N^2}
e^{-\frac{L^2\sigma^2}{2}}f(L^2\sigma^2)\right]^2, 
\label{leading-Wilsonloop}
\eeq
where $f(x)$ is a polynomial of $x$ given by 
\bea
f(x) & \equiv & \sum_{n=1}^{N-1}\sum_{m=0}^{n-1}n!m!
\left[
\left\{\sum_{k=0}^n\frac{1}{(k!)^2(n-k)!}\left(-\frac{x}{2}\right)^k\right\}
\left\{\sum_{l=0}^m\frac{1}{(l!)^2(m-l)!}\left(-\frac{x}{2}\right)^l\right\}\right.\nn 
\\
 & & -\left.\left(\frac{x}{2}\right)^{n-m}
\left\{\sum_{k=0}^m\frac{1}{k!(n-m+k)!(m-k)!}\left(-\frac{x}{2}\right)^k\right\}^2
\right]. 
\eea

\paragraph{First Order Term}

  Next, we compute the first order term $-\bra(S-S_0)\hat{W}(L)\ket_{C,0}$.    
After integrating out the variables other than $X_1$ and $X_2$, we have 
\bea
\lefteqn{-\bra(S-S_0)\hat{W}(L)\ket_{C,0}
= (N^2-1)\left[\frac{\sigma^4}{4g^2}N(2D-3)-1\right]\bra\hat{W}(L)\ket_0} \nn \\
 &  & -\left[\frac{\sigma^2}{2g^2}N(D-2)-\frac{1}{\sigma^2}\right]
\left\{\bra\tr(X_1)^2\hat{W}(L)\ket_0+\bra\tr(X_2)^2\hat{W}(L)\ket_0\right\} \nn \\
 & & -\frac{1}{g^2}\bra(X_1X_1)_{ij}(X_2X_2)_{ji}\hat{W}(L)\ket_0
+\frac{1}{g^2}\bra(X_1)_{ij}(X_1)_{kl}(X_2)_{jk}(X_2)_{li}\hat{W}(L)\ket_0. 
\label{1storder-Wilson0}
\eea
Noting the $SU(N)$-transformation property as in the case of $I(L)_{ijkl}$, 
the calculation of the correlators in the r.h.s. is reduced to 
the following more fundamental quantities: 
\bea
f_1(L)\equiv\bra\frac1N\tr((X_1)^2e^{iLX_1})\frac1N\tr e^{-iLX_1}\ket_0, \nn \\
f_2(L)\equiv\bra\frac1N\tr(X_1)^2\frac1N\tr e^{iLX_1}\frac1N\tr e^{-iLX_1}\ket_0, \nn 
\\
f_3(L)\equiv\bra\frac1N\tr(X_1e^{iLX_1})\frac1N\tr(X_1e^{-iLX_1})\ket_0. 
\eea
The result is 
\bea
\lefteqn{\bra\tr(X_1)^2\hat{W}(L)\ket_0=\bra\tr(X_2)^2\hat{W}(L)\ket_0=
\frac12\sigma^2(N^2-1)} \nn \\
 &+ & \left\{\frac12\sigma^2N^2-\frac{N^3}{N^2-1}f_2(L)\right\}
\left[-1+\frac1N+\frac{2}{N^2}e^{-\frac{L^2\sigma^2}{2}}f(L^2\sigma^2)\right], 
\label{Wilson-S0}
\eea
\bea
\bra(X_1X_1)_{ij}(X_2X_2)_{ji}\hat{W}(L)\ket_0 & = & 
-\frac14\sigma^4\frac{N^2-1}{N}+\sigma^2N^2f_2(L) -\frac{N^3}{N^2-1}(f_2(L))^2 \nn \\
 & & -\frac{4N^3}{(N^2-1)(N^2-4)}(f_1(L)-f_2(L))^2, 
\eea
\bea
\lefteqn{\bra(X_1)_{ij}(X_1)_{kl}(X_2)_{jk}(X_2)_{li}\hat{W}(L)\ket_0=
\frac34\sigma^4\frac{N^2-1}{N(N^2-9)}} \nn \\
 & & -4\sigma^2\frac{N^2+6}{(N^2-4)(N^2-9)}f_1(L) 
-\sigma^2\frac{N^2(N^2-14)}{(N^2-4)(N^2-9)}f_2(L)
-6\sigma^2\frac{1}{N^2-9}f_3(L) \nn \\
 & & +\frac{24N(N^2+1)}{(N^2-1)(N^2-4)(N^2-9)}(f_1(L))^2
-\frac{40N^3}{(N^2-1)(N^2-4)(N^2-9)}f_1(L)f_2(L) \nn \\
 & & +\frac{N^3(N^2+6)}{(N^2-1)(N^2-4)(N^2-9)}(f_2(L))^2 
+\frac{16N(2N^2-3)}{(N^2-1)(N^2-4)(N^2-9)}f_1(L)f_3(L) \nn \\
 & & -\frac{20N^3}{(N^2-1)(N^2-4)(N^2-9)}f_2(L)f_3(L)
+\frac{2N(N^2-3)}{(N^2-1)(N^2-9)}(f_3(L))^2. 
\eea
Now, what is needed for giving the expression of the first order term is 
to know the three quantities $f_1(L)$, $f_2(L)$ and $f_3(L)$. 
We pass to the Gaussian integral of the hermitian matrix $Y$ to evaluate them.  
Going along the same line as before, we obtain the following formulas: 
\bea
\lefteqn{f_1(L)=\frac{\sigma^2}{2}\frac{N^2-1}{N^2}} \nn \\
 & + & \frac{\sigma^2}{2N^2}e^{-\frac{L^2\sigma^2}{2}}
\sum_{(n\neq m)\; n,m=0}^{N-1}\left\{
\left(4n+1-\frac{L^2\sigma^2}{2}-\frac1N\right)F\left(-n,1;\frac{L^2\sigma^2}{2}\right) 
\right. \nn \\
 &   & -4nF\left(-n+1,1;\frac{L^2\sigma^2}{2}\right) 
-2n^2\frac{L^2\sigma^2}{2}F\left(-n+2,3;\frac{L^2\sigma^2}{2}\right)
+4nF\left(-n+2,2;\frac{L^2\sigma^2}{2}\right) \nn \\
 &   & \left.-2nF\left(-n+1,2;\frac{L^2\sigma^2}{2}\right)\right\}
F\left(-m,1;\frac{L^2\sigma^2}{2}\right) \nn \\
 & + & \frac{\sigma^2}{N^2}e^{-\frac{L^2\sigma^2}{2}}\sum_{n=1}^{N-1}\sum_{m=0}^{n-1}
\frac{n!}{m!}\frac{1}{((n-m)!)^2}\left\{\left(\frac{L^2\sigma^2}{2}\right)^{n-m+1}
F\left(-m,n-m+1;\frac{L^2\sigma^2}{2}\right) \right.\nn \\
 &   & -\left(\frac{L^2\sigma^2}{2}\right)^{n-m}\left(2n+2m+1-\frac1N\right)
F\left(-m,n-m+1;\frac{L^2\sigma^2}{2}\right) \nn \\
 &   & +\left(\frac{L^2\sigma^2}{2}\right)^{n-m}\frac{2m}{n-m+1}
F\left(-m+1,n-m+2;\frac{L^2\sigma^2}{2}\right) \nn \\
 &   & \left. +\left(\frac{L^2\sigma^2}{2}\right)^{n-m-1}(n-m)(n-m-1)
F\left(-m,n-m+1;\frac{L^2\sigma^2}{2}\right)\right\} \nn \\
 &   & \times F\left(-m,n-m+1;\frac{L^2\sigma^2}{2}\right), 
\label{f1L}
\eea
\bea
\lefteqn{f_2(L)=\frac{N^2-1}{2N}\sigma^2\left[\frac1N+
\frac{2}{N^2}e^{-\frac{L^2\sigma^2}{2}}f(L^2\sigma^2)\right]} \nn \\
 & + & \frac{\sigma^2}{N^3}e^{-\frac{L^2\sigma^2}{2}}\sum_{n,m=0}^{N-1}\left\{
-\frac{L^2\sigma^2}{2}F\left(-n,1;\frac{L^2\sigma^2}{2}\right)
+2(n-1)\frac{L^2\sigma^2}{2}F\left(-n+2,2;\frac{L^2\sigma^2}{2}\right) \right. \nn \\
 &   & -2F\left(-n+2,2;\frac{L^2\sigma^2}{2}\right)
+n\frac{L^2\sigma^2}{2}F\left(-n+2,3;\frac{L^2\sigma^2}{2}\right) \nn \\
 &   & \left. +2\left(\frac{L^2\sigma^2}{2}+1\right)
F\left(-n+1,1;\frac{L^2\sigma^2}{2}\right)
-4n\frac{L^2\sigma^2}{2}F\left(-n+1,2;\frac{L^2\sigma^2}{2}\right)\right\}
F\left(-m,1;\frac{L^2\sigma^2}{2}\right) \nn \\
 & - & \frac{\sigma^2}{N^3}e^{-\frac{L^2\sigma^2}{2}}\sum_{n=0}^{N-1}\left\{
-\frac{L^2\sigma^2}{2}F\left(-n,1;\frac{L^2\sigma^2}{2}\right)
-2nF\left(-n+1,2;\frac{L^2\sigma^2}{2}\right) 
+2nF\left(-n,1;\frac{L^2\sigma^2}{2}\right) \right. \nn \\
 &   & \left.-2\left(\frac{L^2\sigma^2}{2}+1\right)
F\left(2,2;\frac{L^2\sigma^2}{2}\right)
+2F\left(2,1;\frac{L^2\sigma^2}{2}\right)\right\}
F\left(-n,1;\frac{L^2\sigma^2}{2}\right) \nn \\
 & + & \frac{\sigma^2}{N^3}e^{-\frac{L^2\sigma^2}{2}}2\sum_{n=1}^{N-1}\sum_{m=0}^{n-1}
\frac{n!}{m!}\frac{1}{((n-m)!)^2}\left\{
\left(\frac{L^2\sigma^2}{2}\right)^{n-m+1}F\left(-m,n-m+1;\frac{L^2\sigma^2}{2}\right)
\right. \nn \\
 &   & -\left(\frac{L^2\sigma^2}{2}\right)^{n-m}(n+m)
F\left(-m,n-m+1;\frac{L^2\sigma^2}{2}\right) \nn \\
 &   & +\left(\frac{L^2\sigma^2}{2}\right)^{n-m}\frac{2m}{n-m+1}
F\left(-m+1,n-m+2;\frac{L^2\sigma^2}{2}\right) \nn \\
 &   & \left.+\left(\frac{L^2\sigma^2}{2}\right)^{n-m-1}(n-m)(n-m-1)
F\left(-m,n-m+1;\frac{L^2\sigma^2}{2}\right)\right\} \nn \\
 &   & \times F\left(-m,n-m+1;\frac{L^2\sigma^2}{2}\right) \nn \\
 & + & \frac{\sigma^2}{N^3}e^{-\frac{L^2\sigma^2}{2}}2\sum_{n=2}^{N-1}\sum_{m=1}^{n-1}
\frac{n!}{m!}\frac{1}{(n-m+1)!(n-m-1)!}\left(\frac{L^2\sigma^2}{2}\right)^{n-m} \nn \\
 &   & \times \left\{\left(\frac{L^2\sigma^2}{2}-n-m\right)
F\left(-m,n-m;\frac{L^2\sigma^2}{2}\right)
+(n-m)F\left(-m-1,n-m;\frac{L^2\sigma^2}{2}\right)\right\} \nn \\
 &   & \times F\left(-m+1,n-m+2;\frac{L^2\sigma^2}{2}\right) \nn \\
 & + & \frac{\sigma^2}{N^3}e^{-\frac{L^2\sigma^2}{2}}2\sum_{n=0}^{N-3}
(n+2)(n+1)\frac{L^2\sigma^2}{2}F\left(-n-1,2;\frac{L^2\sigma^2}{2}\right)
F\left(-n,2;\frac{L^2\sigma^2}{2}\right), 
\label{f2L}
\eea
\bea
\lefteqn{f_3(L)=\frac{\sigma^2}{2}\frac{N^2-1}{N^2}} \nn \\
 & + & \frac{\sigma^2}{N^3}e^{-\frac{L^2\sigma^2}{2}}\sum_{n=1}^{N-1}\sum_{m=0}^{n-1}
\left[\frac{L^2\sigma^2}{2}N
\left\{F\left(-n,1;\frac{L^2\sigma^2}{2}\right)
+2nF\left(-n+1,2;\frac{L^2\sigma^2}{2}\right)\right\} \right. \nn \\
 &   & \left.\times\left\{F\left(-m,1;\frac{L^2\sigma^2}{2}\right)
+2mF\left(-m+1,2;\frac{L^2\sigma^2}{2}\right)\right\} 
-F\left(-n,1;\frac{L^2\sigma^2}{2}\right)F\left(-m,1;\frac{L^2\sigma^2}{2}\right)
\right] \nn \\
 & - &  \frac{\sigma^2}{N^3}e^{-\frac{L^2\sigma^2}{2}}\sum_{n=1}^{N-1}\sum_{m=0}^{n-1}
\frac{n!}{m!}\frac{1}{((n-m)!)^2}\left(\frac{L^2\sigma^2}{2}\right)^{n-m-1} \nn \\
 &   & \times \left[N\left\{\left(\frac{L^2\sigma^2}{2}-n-m\right)
F\left(-m,n-m+1;\frac{L^2\sigma^2}{2}\right)+
2mF\left(-m+1,n-m+1;\frac{L^2\sigma^2}{2}\right)\right\}^2 \right. \nn \\
 &   & \left.-\frac{L^2\sigma^2}{2}
\left\{F\left(-m,n-m+1;\frac{L^2\sigma^2}{2}\right)\right\}^2\right]. 
\label{f3L}
\eea
We used some recursion relations among confluent hypergeometric functions 
for later convenience in considering the large $N$ limit.

\paragraph{Gaussian Approximation}

   We consider the above result at the value (\ref{sigmaYM}) to obtain 
the Wilson loop amplitude up to the first order in the Gaussian approximation. 
Since the formula is quite lengthy, we do not write down here directly. 
Instead, we plot the result in the case of various values of $D$ and $N$ below. 
As in the case of the Polyakov loop, the large $N$ limit also yields 
the result remarkably simplified as\footnote{See appendix B for detailed explanation.} 
\bea
\lefteqn{\bra\hat{W}(L)\ket = 
1-\left(\frac{1}{\alpha}(J_1(2\sqrt{\alpha}))^2-1\right)^2} \nn \\
 &   & +\frac{1}{D-1}\frac{2}{\alpha^2}
\left(9(J_1(2\sqrt{\alpha}))^2(J_3(2\sqrt{\alpha}))^2
+8(J_2(2\sqrt{\alpha}))^4\right)+O((S-S_0)^2). 
\label{WilsonlargeN}
\eea
The first line and the first term in the second line in the r.h.s. represent 
the contributions from the leading order term 
and from the first order term, respectively. 
The factor $\frac{1}{D-1}$ in the second line 
means the validity of the interpretation as the mean field approximation. 

   In the $D=4$ case, 
it is interesting to compare our result with the numerical calculation reported in 
ref. \cite{Ambjorn}. Figure 11 in ref. \cite{Ambjorn} is the result 
to be compared with ours. As in the case of the Polyakov loop, we take $Ng^2=48$ 
for the common setting. 
In Fig. 5 we show the result of the leading order alone and that summed 
up to the first order when $N=16$. It can be seen that the result up to the 
first order match well with the numerical result. (See also Fig. 6.) 
We similarly plot the result for the case $N=\infty$ in Fig. 6. 
Comparing this with Fig. 5, 
we expect a good scaling behavior in the region of smaller $L$. 
In fact, Fig. 7 shows the behaviors for $N=16$, 48 and $\infty$ together, from which 
the scaling for smaller $L$ can be observed. 
In particular, the scaling between the cases of $N=48$ and $N=\infty$ is 
too good to distinguish each curve in the region $L<4$. 
These scaling behaviors are in conformity with the numerical result in 
ref. \cite{Ambjorn}. 
\begin{figure}
\epsfxsize=9cm \epsfysize=6cm
\centerline{\epsfbox{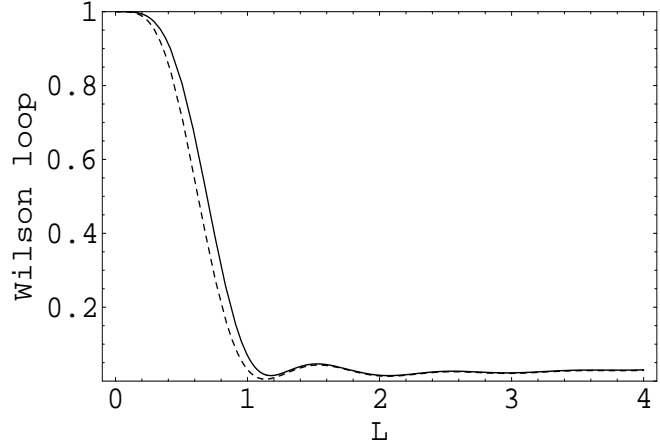}}
\caption{Wilson loop in the case $D=4$ and $N=16$. 
The dotted line shows the contribution from the zeroth order alone, and 
the solid line shows that up to the first order.}
\label{Fig:Wil_d4N16_zero-first}
\end{figure}
\begin{figure}
\epsfxsize=9cm \epsfysize=6cm
\centerline{\epsfbox{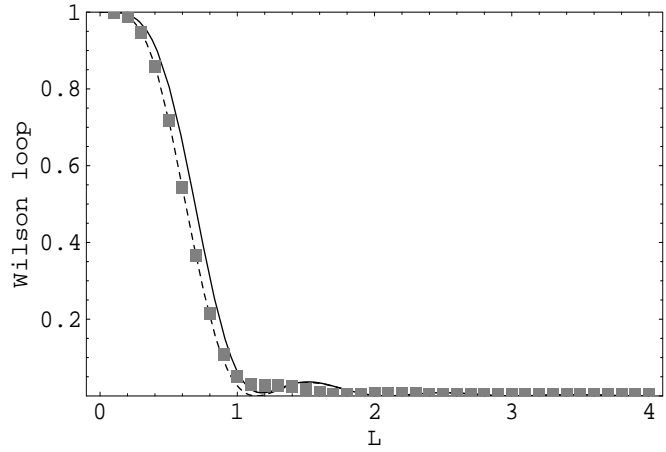}}
\caption{Wilson loop for $D=4$, $N=\infty$ case.
The dotted line shows the contribution from the zeroth order alone, and 
the solid line shows that up to the first order. 
The gray boxes indicate some points of the numerical data 
for the $N=48$ case in Figure 11 in ref. \cite{Ambjorn}.}
\end{figure}                                                                                
\begin{figure}
\epsfxsize=9cm \epsfysize=6cm
\centerline{\epsfbox{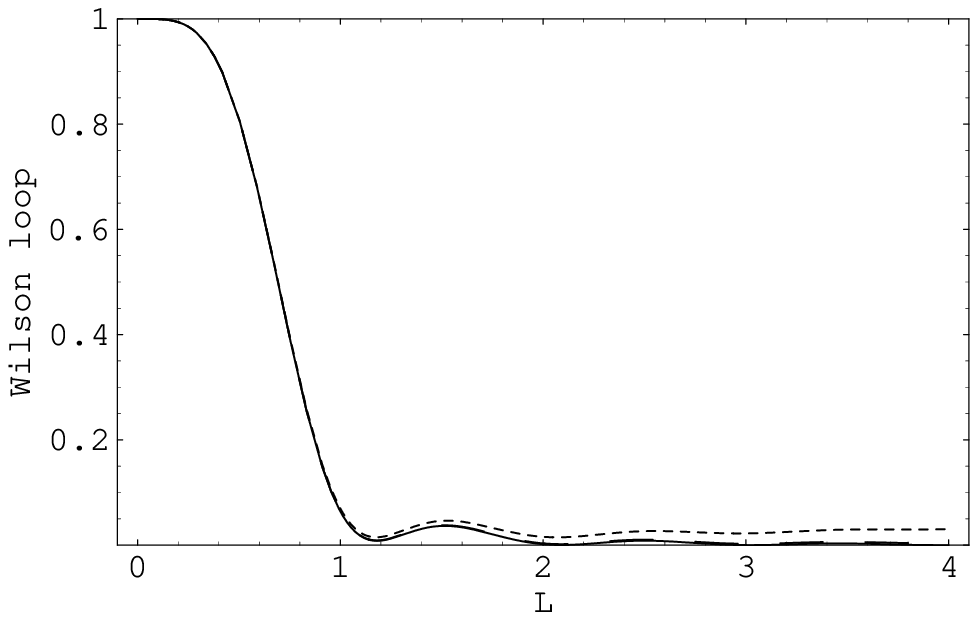}}
\caption{Wilson loops up to the first order 
for $(D,N)=(4,16)$, $(4,48)$ and $(4,\infty)$. 
The dotted line shows the case $(4,16)$, the dashed line $(4,48)$ 
and the solid line $(4,\infty)$.}
\end{figure}

   As in the case of the Polyakov loop, 
we depict the behaviors for $D=10$ and 26 in Fig. 8. 
We can still see quite good scaling behaviors. 
Also, from Fig. 7 and Fig. 8, it can be seen that the curve tends to 
become broader as $D$ increases. The tendency can be understood from the formula 
in the large $N$ limit (\ref{WilsonlargeN}), where $L$ appears only through 
the combination of $\alpha$. 
\begin{figure}
\epsfxsize=9cm \epsfysize=6cm
\centerline{\epsfbox{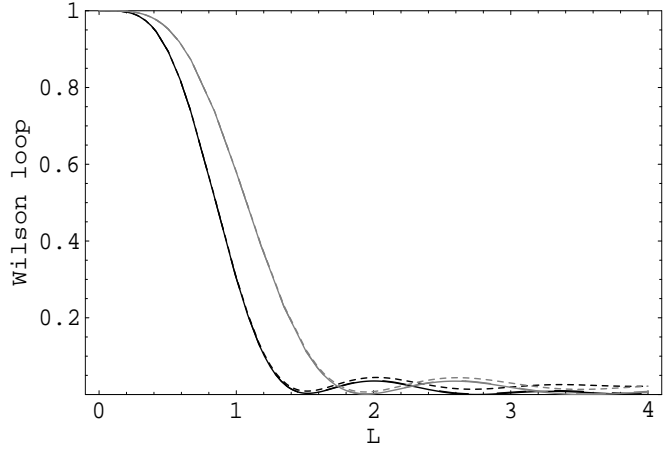}}
\caption{Wilson loops up to the first order 
for $(D,N)=(10,16)$, $(10,48)$, $(10,\infty)$, $(26,16)$, $(26,48)$ and $(26,\infty)$. 
The dotted line shows the case $(10,16)$, the dashed line $(10,48)$, 
the solid line $(10,\infty)$, the gray dotted line $(26,16)$, 
the gray dashed line $(26,48)$, and the gray solid line $(26,\infty)$.}
\end{figure}                                


\section{Reduced Yang-Mills Integrals --- Improved Mean Field Approximation}
\setcounter{equation}{0}
   
   In this section we evaluate the expectation values for the space-time extent, 
the Polyakov loop and the Wilson loop by the method of the improved mean field 
approximation. For the space-time extent, it turns out to give a result closer to 
the exact value than the Gaussian approximation in the case of $N=2$. 
Furthermore, for the Polyakov loop and the Wilson loop, it yields the asymptotic behaviors 
for the large loop. Although we consider only the first few terms in the expansion 
around 
the mean field configuration, as we have done in section 2.2, 
it can be guessed to give the results which reproduce 
the correct behaviors at least qualitatively. 

  Let us consider the improved mean field approximation in the reduced Yang-Mills 
integrals for the operator $\cO$ which is $U(N)$-invariant and 
isotropic in $D$-dimensional space. We start with the mean field action 
\beq
S_M =\frac{1}{2g^2}\sum_{\mu\neq\nu}2\left[-\bra(X_{\mu})_{ij}(X_{\mu})_{kl}\ket^{\cO} 
(X_{\nu})_{jk}(X_{\nu})_{li}
+\bra(X_{\mu})_{ij}(X_{\mu})_{jk}\ket^{\cO}(X_{\nu})_{kl}(X_{\nu})_{li}\right]
+\frac{1}{\lt}. 
\eeq
Since $\cO$ is $U(N)$-invariant, 
\beq
\bra(X_{\mu})_{ij}(X_{\mu})_{kl}\ket^{\cO}
\equiv
\frac{\int(dX)\;e^{-S}(X_{\mu})_{ij}(X_{\mu})_{kl}\cO}{\int(dX)\;e^{-S}\cO}
\propto\delta_{il}\delta_{jk}-\frac{1}{N}\delta_{ij}\delta_{kl}. 
\eeq
Also, due to the isotropic nature, the proportional constant in the above is 
independent 
of $\mu$. 
So the action $S_M$ can be written as the standard form 
$S_M=\frac{1}{\st^2}\tr (X_{\mu})^2+\frac{1}{\lt}$ with 
\beq
\frac{1}{\st^2}=
\frac{1}{g^2}\frac{D-1}{D}\frac{N^2}{N^2-1}\bra\frac1N\tr(X_{\mu})^2\ket^{\cO}. 
\eeq
The self-consistency condition 
$\bra\frac1N\tr(X_{\mu})^2\ket^{\cO}=\bra\frac1N\tr(X_{\mu})^2\ket^{\cO}_M$ reads 
\beq
\frac{1}{\st^2}g^2\frac{D}{D-1}\frac{N^2-1}{N^2}=
\frac{\bra\frac1N\tr(X_{\mu})^2\cO\ket_M}{\bra\cO\ket_M}, 
\label{YM-st}
\eeq
which determines $\st$. 
As before, $\lt$ is given by the formula of the same form as (\ref{example2-lt}) with 
$S'_M=\frac{1}{\st^2}\tr (X_{\mu})^2$.

\subsection{Space-Time Extent}  
   
  Now we consider the case $\cO=\frac1N\tr(X_{\mu})^2$. From (\ref{YM-st}), we have 
\beq
\st^2=\sqrt{\frac{2g^2}{N(D-1)}}\left[1+\frac{2}{(N^2-1)D}\right]^{-1/2}. 
\eeq
Up to the second order with respect to $S-S'_M$, $\lt$ becomes 
\bea 
\frac{1}{\lt} & = & \frac{3\st^4}{8g^2}N(N^2-1)D(D-1)-\frac34(N^2-1)D+
\frac{3\st^4}{2g^2}N(D-1)-\frac32 \nn \\ 
 &   & -\frac{\st^8}{64g^4}N^2(N^2-1)D(D-1)(4D-1)-\frac{3\st^8}{8g^4}N^2D(D-1)
+O((S-S'_M)^3). 
\eea
Following the steps in section 2.2, 
we obtain the expectation value 
\bea
\lefteqn{\bra\frac1N\tr(X_{\mu})^2\ket= \sqrt{\frac{Ng^2}{2(D-1)}}\frac{N^2-1}{N^2}D
\left[1+\frac{2}{(N^2-1)D}\right]^{-\frac14(N^2-1)D-\frac12}}\nn \\
 & &\times\exp\left\{ 
\left[\frac12+\frac{3}{4(D-1)}+\frac{2}{(N^2-1)D}
-\frac{3}{4(D-1)}\frac{1}{(N^2-1)D}+\frac{6}{(N^2-1)^2D^2}\right]\right. \nn \\
 & & \hspace{2cm}\times\left.\left[1+\frac{2}{(N^2-1)D}\right]^{-2}\right\} \nn \\
 &   & \times [1+O((S-S'_M)^3)]. 
\eea
Then, in the case of $N=2$ and $D$ large, it has the following expansion: 
\beq
\bra\frac12\tr(X_{\mu})^2\ket=\frac{3g}{4}\sqrt{D}\left[1+\frac{13}{12D}+O(D^{-2})
+O((S-S'_M)^3)\right]. 
\eeq
Here, it is noted that this result fits the exact result (\ref{N=2exactlargeD}) 
better than the case of the Gaussian approximation (\ref{GaussianlargeD}). 
So, this scheme can be expected to give an improved result also for other $N$.
Moreover, there is another comment. 
If we stopped the computation up to the first order, we would end up with the expansion 
\beq
\bra\frac12\tr(X_{\mu})^2\ket=\frac{3g}{4}\sqrt{D}\left[1+\frac{2}{3D}+O(D^{-2})
+O((S-S'_M)^2)\right]. 
\eeq
Thus, we can see that in the $SU(2)$ case the result approaches closer to the exact 
result as increasing the order of the approximation. 
For general $N$, we can compare with the $1/D$-expansion result (\ref{1/D-expansion}). 
Normalizing the leading term to the unity, we have the following coefficients in the 
$O(D^{-1/2})$-terms: 
\beq
\begin{array}{ll}
2           & :\mbox{Gaussian approximation up to the 2nd order} \\ 
\frac54-\frac{1}{2(N^2-1)} & 
:\mbox{Improved mean field approximation up to the 2nd order} \\
\frac12+\frac{1}{2(N^2-1)} & 
:\mbox{Improved mean field approximation up to the 1st order} \\
\frac76+\frac{1}{N^2-1} & :\mbox{$1/D$-expansion method}, 
\end{array}
\eeq
which exhibits a similar tendency as seen in the $N=2$ case. Thus it can be expected 
that the expansion with respect to $S-S'_M$ based on the improved mean field 
approximation does work for general $N$.

\subsection{Polyakov Loop}

     Here we consider the case 
$\cO=\frac{1}{2D}\sum_{\mu=1}^D\frac{1}{N}\tr(e^{iLX_{\mu}}+e^{-iLX_{\mu}})$, 
whose expectation value is equal to $\bra\hat{P}(L)\ket$ if $O(D)$-symmetry is not 
broken spontaneously. In the case that $N$ and $D$ are finite, it always holds. 
Let us consider this case\footnote{In the bosonic case which we are considering here, 
we can see no evidence of the spontaneous symmetry breaking by the calculation 
based on the $1/D$-expansion method and by the numerical simulation 
as in ref. \cite{HNT}. So we can extend the results here to the $N=\infty$ case. 
We thank J. Nishimura for informing of this fact.}. 
For this $\cO$, the method in the beginning of this section can be used. 
Now, eq. (\ref{YM-st}) leads 
\bea
\frac{1}{\st^2}g^2\frac{D}{D-1}\frac{N^2-1}{N^2}
 & = & -\frac{L^2\st^4}{4}\frac{N+1}{N}
\frac{F\left(1-N,3;\frac{L^2\st^2}{2}\right)}{F\left(1-N,2;\frac{L^2\st^2}{2}\right)}
\nn \\
 &   & +\left(\frac12(N^2-1)D+\frac{L^2\st^2}{4}\frac{N+1}{N}\right)\frac{\st^2}{N}. 
\eea
We are interested in the case of $L^2g$ and $x\equiv L^2\st^2$ large. 
In this case, the ratio between 
the confluent hypergeometric functions is expanded as 
$$
\frac{F\left(1-N,3;\frac{x}{2}\right)}{F\left(1-N,2;\frac{x}{2}\right)}
=\frac{2}{N+1}\left[1-(N-1)\frac{2}{x}-N(N-1)\frac{4}{x^2}+O(x^{-3})\right], 
$$
so the above equation has the form 
\beq
-\frac{g^2L^4}{x^2}\frac{D}{D-1}\frac1N=\frac{1}{4N(N+1)}x-\frac12D-\frac{1}{N+1}
-\frac{2N}{N+1}\frac{1}{x}+O(x^{-2}). 
\label{gapequation-P}
\eeq
We iteratively solve this equation and get the three solutions: 
\bea
x_{\pm} & = & e^{\pm\pi i/3}\left(\frac{4D}{D-1}(N+1)L^4g^2\right)^{1/3}+\frac23N((N+1)D+2)
+O((L\sqrt{g})^{-4/3}), \nn \\
x_0 & = & -\left(\frac{4D}{D-1}(N+1)L^4g^2\right)^{1/3}+\frac23N((N+1)D+2)
+O((L\sqrt{g})^{-4/3}). 
\eea
Corresponding to each $x$, $\lt$ is determined by 
\beq
\frac{1}{\lt}=(N-1)\left\{\frac{1}{4N}x-1+O(x^{-1})+O((S-S'_M)^2)\right\}
\eeq
up to the first order with respect to $S-S'_M$, and 
\bea
\frac{1}{\lt} & = & (N-1)\left\{\frac{1}{2N}x-\frac18(N+1)(N-9)D
-\frac18(N+1)^2\frac{D}{D-1}
-\frac32 \right. \nn \\
 &  & \left. +O(x^{-1})+O((S-S'_M)^3)\right\}
\eea
up to the second order. Eq. (\ref{gapequation-P}) was used for making 
the expressions simpler. It is noted that the situation here is very similar to that in 
section 2.2. Applying the same recipe as in section 2.2, 
we end up with the final expression: 
\bea
\bra\hat{P}(L)\ket & = & C_{(n)}(L^4g^2)^{-\frac{1}{12}(N-1)((N+1)D-4)}
\exp\left[-u_{(n)}\frac{N-1}{N}\left(\frac{4D}{D-1}(N+1)L^4g^2\right)^{1/3}\right] \nn \\
 &  & \times\cos
\left[\sqrt{3}u_{(n)}\frac{N-1}{N}\left(\frac{4D}{D-1}(N+1)L^4g^2\right)^{1/3}
-\frac{\pi}{6}(N-1)((N+1)D+2)\right] \nn \\
 &  & \times [1+O((L^4g^2)^{-1/3})+O((S-S'_M)^{n+1})], 
\label{improved-Polyakov2}
\eea
where up to the first order ($n=1$): $u_{(1)}=\frac14$, 
and up to the second order ($n=2$): $u_{(2)}=\frac38$. 
The overall constants are 
\beas
C_{(1)} & = & 
\frac{(-1)^{N-1}}{2^{N-2}N!}\left(\frac{N}{2}(D-1)\right)^{\frac14(N^2-1)D}
\left(\frac{4D}{D-1}(N+1)\right)^{\frac16(N-1)((N+1)D+2)}  \\
      &   & \times \exp\left[-\frac{7}{12}(N^2-1)D+\frac13(N-1)\right], \\
C_{(2)} & = & 
\frac{(-1)^{N-1}}{2^{N-2}N!}\left(\frac{N}{2}(D-1)\right)^{\frac14(N^2-1)D}
\left(\frac{4D}{D-1}(N+1)\right)^{\frac16(N-1)((N+1)D+2)}  \\
      &   & \times\exp\left[\frac18(N^2-1)(N+1)\frac{D^2}{D-1}
-2(N^2-1)D-\frac{3}{16}(N^2-1)\frac{D}{D-1}+\frac12(N-1)\right]. 
\eeas
 
  At present, we have no result of the computer simulation for the behavior in 
the region of $L^2g$ large. 
However, our result can be expected to represent the behavior correctly 
at least qualitatively. 
It seems interesting to consider our result together with the asymptotic behavior 
of the eigenvalue density $\rho(\lambda)$ (\ref{eigenvalue-density}). 
In fact, $\rho(\lambda)$ can be obtained from $\bra \hat{P}(L)\ket$ via the 
Fourier transformation: 
$$
\rho(\lambda)=\int_{-\infty}^{\infty}\frac{dL}{2\pi}\;
e^{-iL\lambda}\bra \hat{P}(L)\ket. 
$$
If we tried to evaluate the integral in the r.h.s. for $\lambda$ large 
by the saddle point approximation 
using the formula of (\ref{improved-Polyakov2}), 
it would turn out that the width of the Gaussian integral around the saddle point 
becomes infinitely large as $\lambda\limit\infty$. 
It indicates that we should take into account the contribution infinitely far away from 
the saddle point, and thus that in this case the saddle point approximation is inadequate. 
We need the information of $\bra\hat{P}(L)\ket$ for the whole range of $L$ 
to get the asymptotic behavior of $\rho(\lambda)$. 

  Also, we can consider the improvement of the result in the region of 
smaller $L^2g$ in the previous section 
by using this method similarly as in the case of the space-time extent 
in the previous subsection. 
Since it is considered not to change the result essentially, 
we do not present here.

\subsection{Wilson Loop}

  We analyze the expectation value of the Wilson loop for the case of 
the length of the loop $L$ large in the improved mean field approximation. 
Doing the same procedure as before for the operator 
\beq
\cO=\frac{1}{D(D-1)}\sum_{\mu\neq\nu}\frac1N\tr\left(
e^{iLX_{\mu}}e^{iLX_{\nu}}e^{-iLX_{\mu}}e^{-iLX_{\nu}}\right), 
\eeq
we have the following equation for $\st$: 
\beq
\frac{1}{\st^2}g^2\frac{D}{D-1}\frac{N^2-1}{N^2}=
\frac{\bra\frac1N\tr(X_{\mu})^2\hat{W}(L)\ket_M}{\bra\hat{W}(L)\ket_M}, 
\label{st-Wilson}
\eeq
where the r.h.s. can be easily evaluated 
by using the expressions (\ref{leading-Wilsonloop}) and (\ref{Wilson-S0}) replaced 
$\sigma$ with $\st$. When $L^2\st^2$ and $L^2g$ are large, the r.h.s. 
is expanded as 
\bea
\lefteqn{\frac{\bra\frac1N\tr(X_{\mu})^2\hat{W}(L)\ket_M}{\bra\hat{W}(L)\ket_M}
=\frac{D}{2}\st^2\frac{N^2-1}{N}} \nn \\
 & & +4\st^2\frac{N-1}{(N!)^2}e^{-\frac{L^2\st^2}{2}}
\left(\frac{L^2\st^2}{2}\right)^{2N-2}
\left[1+O\left(e^{-\frac{L^2\st^2}{2}}(L^2\st^2)^{2N-3}\right)\right]. 
\eea
So, the solution up to the next-to-leading order becomes 
\beq
\st^2=\sigma^2\left\{1-\frac{4}{(N+1)!(N-1)!}\frac1De^{-\frac{L^2\sigma^2}{2}}
\left(\frac{L^2\sigma^2}{2}\right)^{2N-2}
+O\left(e^{-L^2\sigma^2}(L^2\sigma^2)^{4N-4}\right)\right\}, 
\eeq
with $\sigma^2=\sqrt{\frac{2g^2}{N(D-1)}}$. 
Note that in the $L\limit \infty$ limit 
$\st^2$ coincides to the value of $\sigma^2$ in the Gaussian approximation. 
Up to the first order with respect to $S-S'_M$, $\lt$ is determined as 
\bea
\frac{1}{\lt} & = & \frac{\bra(S-S'_M)\hat{W}(L)\ket_M}{\bra\hat{W}(L)\ket_M} 
+O((S-S'_M)^2)\nn \\
 & = & -(N^2-1)\left\{\frac{D}{4}+\frac{1}{N^2(N+3)}\frac{1}{D-1}\right\} \nn \\
 &   & -\frac{2}{N!(N-2)!}\left\{1+\frac{4}{N^2(N+3)}\frac{1}{D}
-\frac{2(N+1)(N+6)}{N^2(N+2)(N+3)}\frac{1}{D-1}\right\} 
e^{-\frac{L^2\sigma^2}{2}}\left(\frac{L^2\sigma^2}{2}\right)^{2N-2}\nn \\
 &   &
+O\left(e^{-\frac{L^2\sigma^2}{2}}(L^2\sigma^2)^{2N-3}\right)
+O((S-S'_M)^2). 
\eea
Plugging these into the unnormalized expectation value 
$\bra\hat{W}(L)\ket'=(\st^2)^{\frac12(N^2-1)D}e^{-1/\lt}\bra\hat{W}(L)\ket_M,$ 
and dividing it by the partition function up to the first order, 
we end up with the final expression: 
\bea
\lefteqn{\bra\hat{W}(L)\ket=\frac{2}{N+1}
\exp\left[\frac{N^2-1}{N^2(N+3)}\frac{1}{D-1}\right]}
\nn \\
 & \times & \left[1+\frac{1}{N!(N-2)!}\frac{4}{N^2(N+3)}
\left\{\frac{2}{D}-\frac{(N+1)(N+6)}{N+2}\frac{1}{D-1}\right\}
e^{-\frac{L^2\sigma^2}{2}}
\left(\frac{L^2\sigma^2}{2}\right)^{2N-2} \right. \nn \\
 &  &\left. +O\left(e^{-\frac{L^2\sigma^2}{2}}(L^2\sigma^2)^{2N-3}\right)
+O((S-S'_M)^2)\right]. 
\label{Wilson-last}
\eea
There are two comments in order. First, in the above expression, $\bra\hat{W}(L)\ket$ 
has a nonzero limit as $L\limit \infty$, which is $\frac{2}{N+1}
\exp\left[\frac{N^2-1}{N^2(N+3)}\frac{1}{D-1}\right]$. 
The value approaches to zero as increasing $N$. 
Second, since $\frac{2}{D}-\frac{(N+1)(N+6)}{N+2}\frac{1}{D-1}<0$ for $D>1$, 
$\bra\hat{W}(L)\ket$ approaches to the limit from the below. 
In the analysis of the Wilson loop, 
because $\st$ reduces to $\sigma$ in the $L\limit \infty$ limit,  
the result in the Gaussian approximation is considered to 
roughly reproduce the nonzero limit here. 
In fact, in the $L\limit \infty$ limit the result of the Gaussian approximation 
becomes 
\beq
\lim_{L\limit\infty}\left[\bra\hat{W}(L)\ket_0-
\bra (S-S_0)\hat{W}(L)\ket_{C,0}\right]=\frac{2}{N+1}
+\frac{2}{D-1}\frac{N-1}{N^2(N+3)}, 
\eeq
which reproduces the expansion of the exponential up to the first order 
in the limit of (\ref{Wilson-last}). 
As is seen from the above formula, 
this seemingly strange behavior becomes hardly visible as increasing $N$. 
The phenomenon just like this is reported 
in ref. \cite{Krauth}, 
where for $N=2$, 4 and 8 cases in the supersymmetric IKKT integral 
numerical simulations are performed and results similar as ours are obtained 
(in Fig. 3a in ref. \cite{Krauth})\footnote{According to M. Staudacher, 
the authors of ref. \cite{Krauth} have the result of the simulation for 
the Wilson loop also in the bosonic case. It shows the similar behavior as in the 
supersymmetric case \cite{Staudacher}, which match with our result.}. 
There, it is interpreted as a finite $N$ artifact, 
because of the value of the limit decreasing as increasing $N$, 
which is common to our result. 
Also, its dependence on the shape of the contour is reported 
in ref. \cite{Krauth}. 
That is, when changing the contour from the square to a regular polygon,  
the value decreases as increasing the number of edges of the polygon.

   In this method, we can also consider the improvement of the result 
presented in the region of $L^2g$ smaller in the previous section. 
Due to the same reasoning as in the case of the Polyakov loop, we do not give here.


\section{Discussions}
\setcounter{equation}{0}

    Here, we summarize our results and discuss possible future directions 
for concluding this paper. 

  We analyzed the reduced Yang-Mills integrals by using the Gaussian approximation 
and its improved version, after confirming the validity of the schemes in the 
simple example of the $\phi^4$-integral. 
The free energy was evaluated approximately. 
Since the Gaussian approximation can be regarded as the mean field approximation, 
the approximation is expected to become better 
as the space-time dimensionality $D$ increases. 
Actually, for the $SU(2)$ case, where the exact result is known, 
we can make sure this expectation. Also for several lower $N$'s, 
comparing with the numerical result, we found a tendency of better agreement not only 
for larger $D$ but also for larger $N$.  
Next, we investigated the square of the space-time extent 
and the one-point functions for the Polyakov 
loop and the Wilson loop both by the Gaussian approximation and by the improved mean 
field approximation. In the $SU(2)$ case, we saw that 
our results for the space-time extent exhibit good agreement 
with the exact result when $D$ is large. 
In particular, in the improved mean field approximation, 
the situation becomes better than that 
in the Gaussian approximation. 
We saw that it likely holds for general $N$ by comparing with the $1/D$-expansion result. 
We evaluated the Polyakov loop and the Wilson loop by the Gaussian approximation when 
the length of the loop $L$ is smaller, 
and by the improved mean field approximation when $L$ large. 
In the former analysis, we saw that our results nicely fit the numerical results 
in ref. \cite{Ambjorn}. Furthermore, the remarkably simple formulas were obtained 
in the 't Hooft like large $N$ limit. We also observed quite good scaling 
behaviors 
in the region $N\geq 48$, which means that in the scaling region 
the simple formulas represent 
sufficiently well the behaviors in the case $L$ smaller. 
In the latter analysis, the result of the Wilson loop is in conformity with 
the numerical result \cite{Staudacher}, while with respect to the Polyakov loop 
we do not have any results to be compared with ours 
as far as we know. From the analysis for the simple example, however, 
our results can be expected to reproduce the correct behaviors at least qualitatively. 

  For possible future directions, we mainly point out the following two issues. 
One is an extension of this method for supersymmetric systems. 
Because our motivation is to explore approximation schemes trusted nonperturbatively 
in the IKKT integral, it is one of the most important issues. 
As the first step, it will be good to consider $D=4$ supersymmetric case. 
Since in this case there are nice numerical results in ref. \cite{Ambjorn}, it will be a 
good test for the approximation. From the analysis 
for various supersymmetric quantum mechanical systems in the Gaussian approximation 
in refs. \cite{KL,KLL}, it seems to be necessary to consider the Gaussian approximation 
after rewriting the system in terms of the superfields. 
The other is to calculate more general correlators in our framework. 
In order to use our improved mean field formalism for the one-point functions, 
we considered the restricted class of the operators with the isotropic nature. 
Due to this nature, we could use the simple mean field action. 
For considering general operators, 
we will need to start with the mean field action with more general form. 
We hope that the analysis along this line 
gives any insights for studies of the nonperturbative 
aspects of string theory.

\vspace{1cm}


\begin{large}

{\bf Acknowledgements}

\end{large}

\vspace{7mm}

  The authors would like to thank M. Staudacher for informing them of the results 
not reported in literatures and for giving useful comments. 
Also, they wish to thank the authors of ref. \cite{Ambjorn}, 
in particular J. Nishimura and 
W. Bietenholz for comments and sending numerical data in the paper.  
One of them (F.S.)  thanks A. Allahverdyan for explaining about 
the computer systems at Saclay.

\vspace{1cm}


\begin{large}

{\bf Appendix}

\end{large}

\appendix

\renewcommand{\theequation}{\Alph{section}.\arabic{equation}}


\section{Improved Mean Field Analysis in $\phi^6$-integral}

\setcounter{equation}{0}

  Here we present the improved mean field analysis for $\phi^6$-integral 
as another example which holds the assumption mentioned in section 2.2. 
We consider the $\phi^6$-integral defined by the classical action:
$S=\frac{1}{6g^2}\phi^6$. 

\subsection{Saddle Point Approximation}
 
  First, let us evaluate the integral 
$\int_{-\infty}^{\infty}d\phi\;e^{-S}e^{iL\phi}$ 
in the case of $L$ large by the saddle point approximation. 
The solutions for the saddle point equation and the corresponding values of 
$-S+iL\phi$ are 
\beq
\phi_k=e^{\pi i (4k-3)/10}(Lg^2)^{1/5},\hspace{2cm}
(-S+iL\phi)|_{\phi_k}=-\frac{5}{6g^2}e^{2\pi i(k-2)/5}(Lg^2)^{6/5}, 
\label{phi6-saddle-point-value}
\eeq
where $k=1,\cdots, 5$. 
Taking into account regions where the integrand does not blow up as 
$|\phi|\limit\infty$, it turns out that we should deform the integration contour 
so as to pass the three points $\phi_1$, $\phi_2$ and $\phi_3$ 
along each steepest descent direction. 
After the Gaussian integrals, dividing by the partition function 
\beq
Z=\int_{-\infty}^{\infty}d\phi\;e^{-S}=\frac13(6g^2)^{1/6}\Gamma\left(\frac16\right), 
\eeq
the expectation value becomes 
\bea
\lefteqn{\bra e^{iL\phi}\ket = \frac1Z\bra e^{iL\phi}\ket'
\sim \sqrt{\frac{2\pi}{5}}\frac{3}{6^{1/6}}\frac{1}{\Gamma(\frac16)}
\frac{1}{(Lg^{1/3})^{2/5}}\left\{e^{-\frac56(Lg^{1/3})^{6/5}} \frac{}{}\right.}\nn \\ 
 & & \left.+2e^{-\frac56(\cos\frac{2\pi}{5})(Lg^{1/3})^{6/5}}
\cos\left[\frac56(\sin\frac{2\pi}{5})(Lg^{1/3})^{6/5}-\frac{\pi}{5}\right]\right\}. 
\label{phi6-loop}
\eea

\subsection{Improved Mean Field Approximation}

Next, we consider the mean field treatment for this system. 
For the partition function, we start with the following mean field action: 
\beq
S_M=\frac{1}{6g^2}\cdot 15\bra\phi^4\ket \phi^2+\frac{1}{\lambda}
=\frac{1}{2\sigma^2}\phi^2+\frac{1}{\lambda}. 
\eeq
The self-consistency condition $\bra\phi^4\ket=\bra\phi^4\ket_M$ 
determines $\sigma$ as $\sigma^2=\left(\frac{g^2}{15}\right)^{1/3}.$
Also, $\lambda$ is given by the same form as (\ref{example-lambda}). 
Calculation of the first few terms in the mean field expansion leads the expression of 
the free energy: 
\beq
F=-\ln Z=-\frac16\ln g^2-\frac12\ln \frac{2\pi}{15^{1/3}}-\frac13-\frac{17}{45}
+O(\bra(S-S'_M)^3\ket_{C,M}), 
\eeq
where the third and the fourth term in the r.h.s. represent the contribution from 
$\bra S-S'_M\ket_M$ and from $\bra(S-S'_M)^2\ket_{C,M}$, respectively. 

   Now let us discuss the improved version of the mean field approximation 
for the unnormalized expectation value $\bra e^{iL\phi}\ket'$. 
Starting with the mean field action 
\beq
S_M=\frac{1}{6g^2}\cdot 15\bra\phi^4\ket^{\cO} \phi^2+\frac{1}{\lt}
=\frac{1}{2\st^2}\phi^2+\frac{1}{\lt} 
\eeq
for the case of $\cO=e^{iL\phi}$, the self-consistency condition 
$\bra\phi^4\ket^{\cO}=\bra\phi^4\ket^{\cO}_M$ reads 
\beq
x^5=\frac15L^6g^2+6x^4-3x^3
\eeq
where we put $x=L^2\st^2$. Since we are interested in the case that 
$x$ and $L^6g^2$ are large, this equation can be solved iteratively. 
Up to the next-to-leading order, we have the five solutions: 
\beq
x_k =  e^{2\pi i(k-1)/5}\left(\frac15L^6g^2\right)^{1/5}
+\frac65+O((L^6g^2)^{-1/5}), \hspace{2cm} (k=1,\cdots, 5). 
\eeq
$\lt$ is given by eq. (\ref{example2-lt}) corresponding to each solution of $x$. 
After some calculation with respect to the first few terms, 
we obtain the expression of $\lt$ as 
\beq
\frac{1}{\lt}=\frac{7}{15}x-\frac{1}{5}+O(x^{-1})+O((S-S'_M)^2) 
\eeq
up to the first order, and as 
\beq
\frac{1}{\lt}=\frac{59}{75}x+\frac{11}{25}+O(x^{-1})+O((S-S'_M)^3) 
\eeq
up to the second order. 

   The unnormalized expectation value is expressed by the same form 
as (\ref{example2-loop}) also in this case. 
Now, we compare the exponentiated term $-\frac12x-\frac{1}{\lt}$ 
in eq. (\ref{example2-loop}) at each solution $(x, \lt)$ with 
eq. (\ref{phi6-saddle-point-value}) in the saddle point calculation. 
Then, the following precise correspondence similar as in the $\phi^4$ case 
can be found: 
\beq
(x_1,\lt_1)\leftrightarrow \phi_2, \hspace{5mm} 
(x_2,\lt_2)\leftrightarrow \phi_3, \hspace{5mm}
(x_3,\lt_3)\leftrightarrow \phi_4, \hspace{5mm} 
(x_4,\lt_4)\leftrightarrow \phi_5, \hspace{5mm}
(x_5,\lt_5)\leftrightarrow \phi_1. 
\eeq
Thus, we proceed along the same strategy as in the $\phi^4$ case. 
The solutions $(x_3,\lt_3)$ and $(x_4,\lt_4)$ lead 
an unphysical result blowing up as $L\limit \infty$, 
so we discard them. 
Let us combine the contribution from the other solutions with the weight unity. 
Dividing by the partition function, we eventually obtain 
\bea
\bra e^{iL\phi}\ket & = & \frac{15^{1/6}}{5^{1/10}}e^{-u_n}
\frac{1}{(Lg^{1/3})^{2/5}}
\left\{e^{-v_n\cdot 5^{-1/5}(Lg^{1/3})^{6/5}} \frac{}{}\right. \nn \\
 & & \left.+2e^{-v_n\cdot 5^{-1/5}(\cos\frac{2\pi}{5})(Lg^{1/3})^{6/5}}
\cos\left[v_n\cdot 5^{-1/5}(\sin\frac{2\pi}{5})(Lg^{1/3})^{6/5}
-\frac{\pi}{5}\right]\right\} \nn \\
 & &\times [1+O((Lg^{1/3})^{-6/5})+O((S-S'_M)^{n+1})],  
\eea
where up to the first order ($n=1$): $u_1=\frac{97}{75}$, $v_1=\frac{29}{30}$, 
and up to the second order ($n=2$): $u_2=\frac{3032}{1125}$, $v_2=\frac{193}{150}$.  
  
  Let us compare these with the saddle point result (\ref{phi6-loop}). 
The constant factor in front of the whole expression does not exhibit a good 
result as long as looking at the first two orders: 
\beq
\begin{array}{ll}
\frac{15^{1/6}}{5^{1/10}}e^{-\frac{97}{75}}=0.36680\cdots 
&  \mbox{up to the 1st order} \\
\frac{15^{1/6}}{5^{1/10}}e^{-\frac{3032}{1125}}=0.09029\cdots 
&  \mbox{up to the 2nd order} \\
\sqrt{\frac{2\pi}{5}}\frac{3}{6^{1/6}}\frac{1}{\Gamma(\frac{1}{6})}=0.44819\cdots 
&  \mbox{saddle point method}. 
\end{array}
\eeq
However, the coefficient of the exponential decay or the oscillation, which 
plays more important role with respect to the qualitative behavior, 
approaches closer to 
the saddle point result as increasing the precision from the first order to the 
second order: 
\beq
\begin{array}{ll}
\frac{29}{30}\cdot 5^{-1/5}=0.70062\cdots 
&  \mbox{up to the 1st order} \\
\frac{193}{150}\cdot 5^{-1/5}=0.93254\cdots 
&  \mbox{up to the 2nd order} \\
\frac56=0.83333\cdots 
&  \mbox{saddle point method}. 
\end{array}
\eeq


\section{Large $N$ Limit in Polyakov loop and Wilson Loop}

\setcounter{equation}{0}
 
   Here, we give some detailed explanation about the derivation of 
(\ref{PolyakovlargeN}) and (\ref{WilsonlargeN}).  

\subsection{Polyakov Loop}

   First, let us start with the Polyakov loop. 
Eq. (\ref{0thPolyakov}) can be rewritten as 
\bea
\lefteqn{\bra\hat{P}(L)\ket_0 = e^{\frac{\alpha}{2N}\frac{N+1}{N}}
F\left(N+1,2;-\frac{\alpha}{N}\right)} \nn \\
 & = & e^{\frac{\alpha}{2N}\frac{N+1}{N}}\sum_{n=0}^{\infty}
\frac{1}{(n+1)!}\left(1+\frac1N\right)\left(\frac12+\frac1N\right)
\cdots\left(\frac1n+\frac1N\right)(-\alpha)^n,
\eea
with $\alpha=\frac{L^2G}{\sqrt{2(D-1)}}$. 
We consider the $N\limit \infty$ limit keeping $G=\sqrt{N}\,g$ fixed. 
Because the series in the r.h.s. converges uniformly with respect to $N$, 
the order of the limit and the summation $\sum_{n=0}^{\infty}$ can be 
changed. So we obtain 
\beq
\lim_{N\limit\infty}\bra\hat{P}(L)\ket_0= 
\sum_{n=0}^{\infty}\frac{1}{n!(n+1)!}(-\alpha)^n
=\frac{1}{\sqrt{\alpha}}J_1(2\sqrt{\alpha}), 
\eeq
which is the first term in (\ref{PolyakovlargeN}). 

  In the second order result (\ref{2ndPolyakov2}), 
after picking up $O(N^0)$-terms dominant in the large $N$ limit, 
we have 
\bea
\lefteqn{\frac12\bra(S-S_0)^2\hat{P}(L)\ket_{C,0}
=e^{\frac{\alpha}{2N}\frac{N+1}{N}}}\nn \\
 & \times & \left\{\frac{N-3}{8N^5}\frac{\alpha^2}{D-1} 
\left[\left(\begin{array}{c} N \\ 4\end{array}\right)
F\left(N+1,5;-\frac{\alpha}{N}\right)
+\left(\begin{array}{c} N+1 \\ 4\end{array}\right)
F\left(N+2,5;-\frac{\alpha}{N}\right)\right] \right.\nn \\
 &  & +\frac{N+3}{8N^5}\frac{\alpha^2}{D-1}
\left[\left(\begin{array}{c} N+2 \\ 4\end{array}\right)
F\left(N+3,5;-\frac{\alpha}{N}\right)
+\left(\begin{array}{c} N+3 \\ 4\end{array}\right)
F\left(N+4,5;-\frac{\alpha}{N}\right)\right] \nn \\
 & & -\frac{3N-1}{4N^3}\frac{\alpha}{D-1}
\left(\begin{array}{c} N\\2\end{array}\right)
F\left(N+1,3;-\frac{\alpha}{N}\right) \nn \\
 & & \left. -\frac{3N+1}{4N^3}\frac{\alpha}{D-1}
\left(\begin{array}{c} N+1\\2\end{array}\right)
F\left(N+2,3;-\frac{\alpha}{N}\right)\right\}
+O(N^{-2}). 
\eea
Considering the limit of each confluent hypergeometric function 
as above, we arrive at the formula 
\beq
\lim_{N\limit\infty}\frac12\bra(S-S_0)^2\hat{P}(L)\ket_{C,0}
=\frac{1}{2(D-1)}[J_4(2\sqrt{\alpha})-3J_2(2\sqrt{\alpha})]. 
\eeq
This gives the second term in (\ref{PolyakovlargeN}).

\subsection{Wilson Loop}

   Next, let us argue about the Wilson loop. 
We start with considering the limit with respect to the function $f(L^2\sigma^2)$ 
appearing in the expression of the leading order 
term (\ref{leading-Wilsonloop}): 
\bea
\lefteqn{\frac{1}{N^2}f(L^2\sigma^2)= e^{2\alpha/N}\frac{1}{N^2}
\sum_{n=1}^{N-1}\sum_{m=0}^{n-1}
F\left(n+1,1;-\frac{\alpha}{N}\right)F\left(m+1,1;-\frac{\alpha}{N}\right)} 
\nn \\
 & & -e^{2\alpha/N}\frac{1}{N^2}
\sum_{n=1}^{N-1}\sum_{m=0}^{n-1}
\left(\frac{\alpha}{N}\right)^{n-m}\frac{n!}{m!((n-m)!)^2}
\left(F\left(n+1,n-m+1;-\frac{\alpha}{N}\right)\right)^2.
\eea
We put $n=Nx$ and $m=Ny$. In the $N\limit\infty$ limit, $x$ and $y$ become 
running over the interval $0< x<1$, $0\leq y<1$ continuously. 
Then, the first term is expressed as 
\bea
\int^1_0dx\int^x_0dy\;J_0(2\sqrt{x\alpha})J_0(2\sqrt{y\alpha})
=\frac12\left(\int^1_0dx\;J_0(2\sqrt{x\alpha})\right)^2
=\frac{1}{2\alpha}(J_1(2\sqrt{\alpha}))^2.
\eea
It is seen that the second term can be neglected in the limit, from the following 
consideration. First, we separate the second term into the two parts. 
One is the contribution from the region $n-m=O(N^a)$ $(0<a\leq 1)$, 
and the other is that from $n-m=c=O(N^0)$. 
With respect to the first part, the case of $n=O(N)$ is dominant, and 
then $\frac{n!}{m!}\leq {\rm (const)}N^{bN^a}$, where we wrote 
$n-m=bN^a$ with $b$ being a $O(1)$-constant. 
Also
$$
\left(\frac{\alpha}{N}\right)^{n-m}\frac{n!}{m!((n-m)!)^2}
\leq {\rm (const)}\left(\frac{e^2\alpha}{b^2N^{2a}}\right)^{bN^a}\frac{1}{N^{a}}, 
$$
which is negligible in the limit $N\limit\infty$ for a fixed arbitrary $\alpha$. 
Thus we can neglect the first part. 
Next, let us consider the second part. 
Noting that $n$ and $m$ run satisfying $n-m=c=O(N^0)$, 
powercounting leads 
$
|\mbox{(the second part)}|\leq {\rm (const)}\frac{1}{N^2}\cdot N$. 
So we can also neglect this contribution when $N\limit\infty$. 

  Thus, the first term alone survives in the limit, and it yields the result 
\beq
\lim_{N\limit \infty}\frac{1}{N^2}f(L^2\sigma^2)
=\frac{1}{2\alpha}(J_1(2\sqrt{\alpha}))^2. 
\eeq
As the consequence, we obtain the limit of the leading order term 
\beq
\lim_{N\limit\infty}\bra\hat{W}(L)\ket_0=
1-\left[\frac{1}{\alpha}(J_1(2\sqrt{\alpha}))^2-1\right]^2. 
\label{leadingWilson2}
\eeq

\paragraph{First Order Term} 
   
For considering the large $N$ limit of the first order term, 
it is convenient to rewrite the r.h.s. of eq. (\ref{1storder-Wilson0}) 
in terms of connected correlators. Namely, 
\bea
\lefteqn{-\bra(S-S_0)\hat{W}(L)\ket_{C,0}
=\frac{2N}{\sqrt{2G^2(D-1)}}\bra\tr(X_1)^2\hat{W}(L)\ket_{C,0}}\nn \\
 & & -\frac{N}{G^2}\bra\tr(X_1X_1X_2X_2)\hat{W}(L)\ket_{C,0}
+\frac{N}{G^2}\bra\tr(X_1X_2X_1X_2)\hat{W}(L)\ket_{C,0}. 
\label{1storder-Wilson1}
\eea
Each connected correlator is given as follows: 
\beas
 & &\bra\tr(X_1)^2\hat{W}(L)\ket_{C,0}=
-\frac{N^3}{N^2-1}\tilde{f}_2(L)\left[-1+\frac{1}{N}
+\frac{2}{N^2}e^{-\frac{L^2\sigma^2}{2}}f(L^2\sigma^2)\right],  \\
 & & \bra\tr(X_1X_1X_2X_2)\hat{W}(L)\ket_{C,0}=
\sigma^2N^2\tilde{f}_2(L)
-\frac{N^3}{N^2-1}\tilde{f}_2(L)\left\{\tilde{f}_2(L)
+\frac{N^2-1}{N}\sigma^2u(L)\right\} \\
 & & \hspace{1cm}-\frac{4N^3}{(N^2-1)(N^2-4)}\left\{f_1(L)-\tilde{f}_2(L)
-\frac{N^2-1}{2N}\sigma^2u(L)\right\}^2, 
\eeas
\bea
\lefteqn{\bra\tr(X_1X_2X_1X_2)\hat{W}(L)\ket_{C,0}
=\frac{\sigma^4}{2}\frac{N^2+3}{N(N^2-9)}
-4\sigma^2\frac{N^2+6}{(N^2-4)(N^2-9)}f_1(L)} \nn \\
 & & -\sigma^2\frac{N^2(N^2-14)}{(N^2-4)(N^2-9)}\tilde{f}_2(L)
+\sigma^4\frac{N(N^2+11)}{(N^2-4)(N^2-9)}u(L) 
-6\sigma^2\frac{1}{N^2-9}f_3(L) \nn \\
 & & -\frac{40N^3}{(N^2-1)(N^2-4)(N^2-9)}f_1(L)\left\{\tilde{f}_2(L)
+\frac{N^2-1}{2N}\sigma^2u(L)\right\} \nn \\
 & & +\frac{N^3(N^2+6)}{(N^2-1)(N^2-4)(N^2-9)}\tilde{f}_2(L)\left\{
\tilde{f}_2(L)+\frac{N^2-1}{N}\sigma^2u(L)\right\} \nn \\
 & & -\frac{20N^3}{(N^2-1)(N^2-4)(N^2-9)}f_3(L)\left\{\tilde{f}_2(L)
+\frac{N^2-1}{2N}\sigma^2u(L)\right\}\nn \\
 & & +\frac32\sigma^4\frac{N(3N^2-7)}{(N^2-4)(N^2-9)}(u(L))^2
 +\frac{24N(N^2+1)}{(N^2-1)(N^2-4)(N^2-9)}(f_1(L))^2 \nn \\
 & & +\frac{16N(2N^2-3)}{(N^2-1)(N^2-4)(N^2-9)}f_1(L)f_3(L)
+\frac{2N(N^2-3)}{(N^2-1)(N^2-9)}(f_3(L))^2, 
\label{X1X2X1X2-Wilson0}
\eea
where 
\bea
\tilde{f}_2(L) & \equiv &  f_2(L)-\frac{N^2-1}{2N}\sigma^2\left[
\frac{1}{N}+\frac{2}{N^2}e^{-\frac{L^2\sigma^2}{2}}f(L^2\sigma^2)\right], \\
u(L) & \equiv & \frac1N+\frac{2}{N^2}e^{-\frac{L^2\sigma^2}{2}}f(L^2\sigma^2).
\eea

   Now, let us consider the limit of the basic quantities 
$f_1(L)$, $f_3(L)$ and $N^2\tilde{f}_2(L)$.
As the result of the argument similar as in the leading order case, 
for $f_1(L)$ we find that only the second term in eq. (\ref{f1L}) 
is relevant, and that it gives the limit 
\beq
\lim_{N\limit\infty}f_1(L)=\frac{4}{L^2}J_1(2\sqrt{\alpha})
\left[J_1(2\sqrt{\alpha})-\frac{3}{2\sqrt{\alpha}}J_2(2\sqrt{\alpha})\right].
\eeq
Also, for $f_3(L)$, the term containing the summation of the product of 
$2nF(-n+1,2;\frac{L^2\sigma^2}{2})$ and $2mF(-m,1;\frac{L^2\sigma^2}{2})$ 
in eq. (\ref{f3L}) alone becomes relevant. The result is 
\beq
\lim_{N\limit\infty}f_3(L)
=\frac{4\alpha}{L^2}\left(\int^1_0 dx\sqrt{x}J_1(2\sqrt{\alpha})\right)^2
=\frac{4}{L^2}(J_2(2\sqrt{\alpha}))^2. 
\eeq
For $N^2\tilde{f}_2(L)$, we go along the same line. 
With respect to the second, third and last terms in eq. (\ref{f2L}), 
it can be done easily: 
\bea
 & & \lim_{N\limit \infty}N^2
(\mbox{the 2nd and 3rd terms in eq. (\ref{f2L})})
 = -\frac{4\alpha}{3L^2}\left[(J_1(2\sqrt{\alpha}))^2
+(J_2(2\sqrt{\alpha}))^2\right], \nn \\
 & & \lim_{N\limit \infty}N^2(\mbox{the last term in eq. (\ref{f2L})})
=\frac{1}{3L^2}\left[(J_1(2\sqrt{\alpha}))^2+(J_2(2\sqrt{\alpha}))^2\right]. 
\label{lastinf2L}
\eea
In the fourth and fifth terms in eq. (\ref{f2L}), the dominant contribution 
comes from the case of $n-m=c=O(N^0)$. 
\bea
\lefteqn{\lim_{N\limit \infty}N^2(\mbox{the fourth term in eq. (\ref{f2L})})} 
\nn \\  
 & = & -\frac{8\alpha}{L^2}\sum_{c=1}^{\infty}\int^1_0dx\; x
(J_c(2\sqrt{x\alpha}))^2 
+\frac{8\sqrt{\alpha}}{L^2}\sum_{c=1}^{\infty}\int^1_0dx \sqrt{x}
J_c(2\sqrt{x\alpha})J_{c+1}(2\sqrt{x\alpha}) \nn \\
 & & +\frac{4}{L^2}\sum_{c=1}^{\infty}(c^2-c)\int^1_0dx\;(J_c(2\sqrt{x\alpha}))^2
\nn \\
 & = & \frac{1}{3L^2}\left[4\alpha (J_0(2\sqrt{\alpha}))^2+
4(\alpha-2)(J_1(2\sqrt{\alpha}))^2
+4\sqrt{\alpha} J_0(2\sqrt{\alpha})J_1(2\sqrt{\alpha})\right], 
\label{fourthinf2L}
\eea
where in the last step we used the identity $\sum_{n\in {\bf Z}}(J_n(z))^2=1$ 
and some recursion relations among Bessel functions. Similarly, 
\bea
\lefteqn{\lim_{N\limit \infty}N^2(\mbox{the fifth term in eq. (\ref{f2L})})} 
\nn \\  
 & = & -\frac{8\alpha}{L^2}\sum_{c=1}^{\infty}\int^1_0dx\; x
J_{c-1}(2\sqrt{x\alpha})J_{c+1}(2\sqrt{x\alpha}) 
= -\frac{4\alpha}{L^2}\int^1_0dx\; x(J_1(2\sqrt{x\alpha}))^2 \nn \\
 & = & -\frac{1}{3L^2}\left[(J_1(2\sqrt{\alpha}))^2
+(J_2(2\sqrt{\alpha}))^2\right].
\label{fifthinf2L}
\eea
Putting eqs. (\ref{lastinf2L}), (\ref{fourthinf2L}) and 
(\ref{fifthinf2L}) together, we obtain the simple formula 
\beq
\lim_{N\limit\infty}N^2\tilde{f}_2(L)=-\frac{4\sqrt{\alpha}}{L^2}
J_1(2\sqrt{\alpha})J_2(2\sqrt{\alpha}). 
\eeq

  Now we can write down the three connected correlators appearing in 
the r.h.s. of eq. (\ref{1storder-Wilson1}). 
The last equation in (\ref{X1X2X1X2-Wilson0}) seems to be complicated. 
However, since as the result of powercounting 
it is only the last term in this equation 
that gives the leading contribution in the large $N$ 
limit, the expression becomes considerably simple. The result is as follows: 
\bea
 & & \lim_{N\limit\infty} N\bra\tr(X_1)^2\hat{W}(L)\ket_{C,0}
=\frac{4\sqrt{\alpha}}{L^2}J_1(2\sqrt{\alpha})J_2(2\sqrt{\alpha})
\left[\frac{1}{\alpha}(J_1(2\sqrt{\alpha}))^2-1\right], \nn \\
 & & \lim_{N\limit\infty}N\bra\tr(X_1X_1X_2X_2)\hat{W}(L)\ket_{C,0}
=-\frac{8}{L^4}\alpha^{3/2}J_1(2\sqrt{\alpha})J_2(2\sqrt{\alpha}) \nn \\
 & & \hspace{1cm}-\frac{36}{L^4}(J_1(2\sqrt{\alpha}))^2(J_3(2\sqrt{\alpha}))^2
+\frac{8}{L^4}\sqrt{\alpha}(J_1(2\sqrt{\alpha}))^3J_2(2\sqrt{\alpha}), \nn \\
 & &  \lim_{N\limit\infty}N\bra\tr(X_1X_2X_1X_2)\hat{W}(L)\ket_{C,0}
=\frac{32}{L^4}(J_2(2\sqrt{\alpha}))^4. 
\eea
Plugging these into (\ref{1storder-Wilson1}), 
we end up with the final expression: 
\beq
-\bra(S-S_0)\hat{W}(L)\ket_{C,0}=\frac{1}{D-1}\frac{2}{\alpha^2}
\left[9(J_1(2\sqrt{\alpha}))^2
(J_3(2\sqrt{\alpha}))^2+8(J_2(2\sqrt{\alpha}))^4\right]. 
\label{firstWilson2}
\eeq

   Eqs. (\ref{leadingWilson2}) and (\ref{firstWilson2}) give 
the r.h.s. of eq. (\ref{WilsonlargeN}). Thus we complete the derivation.

\vspace{1cm}


\end{document}